\documentclass[12pt]{article}

\usepackage[dvips]{graphicx}
\usepackage{amsmath}
\usepackage{amsfonts}
\usepackage{amssymb}
\usepackage{hyperref}
\usepackage{color}
\usepackage[small]{caption}

\numberwithin{equation}{section}

\newcommand{\be}{\begin{equation}}
\newcommand{\ee}{\end{equation}}
\newcommand{\bea}{\begin{eqnarray}}
\newcommand{\eea}{\end{eqnarray}}

\newcommand{\nn}{\nonumber}

\newcommand{\kleft}{\overleftarrow{k}}
\newcommand{\kright}{\overrightarrow{k}}

\newcommand{\pade}{Pad\'{e} }
\newcommand{\Q}{Q }

\begin{document}

\begin{flushright}
\small
DESY 13-161 \\[1ex]
CERN-PH-TH/2013-203
\end{flushright}

\vskip 22pt

\begin{center}
{\bf \LARGE {
Cosmological perturbation theory\\[0.75ex] at three-loop order
 }}
\end{center}

\vskip 12pt

\begin{center}
\small
{\bf Diego Blas$^{a}$, Mathias Garny$^{b}$, Thomas Konstandin$^{b}$ } \\
\vskip 3pt
{\em $^a$ CERN, Theory Division, 1211 Geneva, Switzerland } \\
{\em $^b$ DESY, Notkestr.~85, 22607 Hamburg, Germany } \\
\end{center}

\vskip 20pt

\begin{abstract}
\vskip 3pt
\noindent
We analyze the dark matter power spectrum at three-loop order in standard perturbation theory of large scale structure. 
We observe that at late times the
loop expansion does not converge even for large scales (small momenta) well within the linear regime,
but exhibits properties compatible with an asymptotic series. 
We propose a technique to restore the convergence in the limit of small momentum,
and use it to obtain a perturbative expansion with improved convergence
for momenta in the range where baryonic acoustic oscillations are present. 
Our numerical three-loop results are compared with data from N-body simulations 
at different redshifts, and we find good agreement within this range.  
\end{abstract}

\newpage

\section{Introduction\label{sec:intro}}

Analytic techniques in cosmological perturbation 
theory experienced a renaissance in the last decade. The main driver of this development was the observation in refs.~\cite{Crocce:2005xy} and \cite{Crocce:2005xz} that 
large perturbative contributions arising from soft effects can be resummed in standard perturbation theory. 
This result lead to a reorganization of standard perturbation theory (SPT) in terms of 
multi-point correlators, known as Renormalized Perturbation Theory (RPT)~\cite{Crocce:2005xy}.
Especially striking is this result for the propagator where the resummed result agrees very well with the measurements in N-body simulations in contrast to SPT.
Motivated by the success of RPT, a plethora of resummation schemes has been invented, mostly with the goal to resum large soft effects (for a collection of methods see~\cite{Nishimichi:2008ry, Carlson:2009it}). 

However, for equal-time correlators as for example the power spectrum, enhancement from pure soft effects 
should be absent due to Galilean invariance~\cite{Scoccimarro:1995if, Kehagias:2013yd, Peloso:2013zw}. 
This was recently demonstrated explicitly in \cite{Blas:2013bpa} at any order in perturbation theory.
Thus, the breakdown of SPT in the description of the power spectrum at late times does not seem  to be related to soft effects (see also refs.~\cite{Anselmi:2012cn, Sugiyama:2013pwa, Sugiyama:2013gza}). Understanding  the actual reason behind this failure and using it to devise better approaches remains an open question.
 It was also recently realized that the  aforementioned cancellation of the soft enhancements can be made manifest by a 
  judicious symmetrization of the corresponding SPT integrands before any integration is performed~\cite{Blas:2013bpa, Carrasco:2013sva}. This facilitates the numerical evaluation of higher loop contributions that are using Monte-Carlo integration techniques. 
Obtaining the cancellation between different contributions after integration 
would be challenging with these integration techniques.

Most of the previous work on the understanding of the non-linear behavior of cosmological perturbation theory has been focused on short scales, where the higher order corrections (computed up to two-loops) surpass the linear predictions (see e.g.~\cite{Crocce:2005xy,Carlson:2009it,Taruya:2012ut}). 
In this work, we are interested in the opposite regime of the power spectrum, namely small momenta. 
In this case, SPT (and its different extensions) is expected to 
converge well rather independently from the redshift \cite{Bernardeau:2001qr, Valageas:2013hxa}. In the first part 
(Sections \ref{sec:eval} and \ref{sec:3Lresults}) we show that this 
is in fact not true. Even though the linear contribution to the power spectrum dominates in the limit $k \to 0$,
the three-loop contribution surpasses the two-loop and even the one-loop 
contribution at late times ($z \sim 0$) and the convergence of the SPT series is questionable. Subsequently (in Section \ref{sec:Pade}), 
we resum the SPT result by means of \pade approximants using as guidance the 
asymptotic behavior in the small momentum regime found in~\cite{Blas:2013bpa}. We also extend these methods
to describe the power spectrum at slightly larger momentum in this Section.
Finally, we conclude in Section~\ref{sec:concl}. Some technical details about our numerical procedure and the
Pad\'e approximation can be found in the Appendices.

\section{Evaluation of the power spectrum\label{sec:eval}}

\subsection{Formalism}

Our results are presented for an idealized case of Standard Perturbation Theory (SPT). This
 assumes that the physics of interest is well described by considering the first two moments
of the Vlasov equation\footnote{The validity of this approach  has been recently challenged in \cite{Carrasco:2012cv,Hertzberg:2012qn}.}, the density contrast $\delta$ and the velocity field \cite{Bernardeau:2001qr,Peebles:Book,Pueblas:2008uv,Pietroni:2011iz}. For the latter we will assume
that the fluid is irrotational and can be represented by a scalar $\theta$.  
We consider an Einstein-de Sitter Universe without dark energy. However, as initial conditions we use 
the dark matter spectrum corresponding to the realistic cosmological parameters as obtained by CAMB \cite{Lewis:1999bs}
and also employ the appropriate linear growth factor $D_+(z)$ for the SPT expansion. We focus on the growing mode and Gaussian initial conditions. For convenience, we work with the two-components field
$\Psi_a$ with $\Psi_1\equiv \delta$ and $\Psi_2\equiv -\theta/\mathcal H$. The rest of our notation  is the one from~\cite{Blas:2013bpa}.
In particular, three-momenta will be denoted by single letters, e.g. $k$.

In SPT, the calculation of cosmological observables is organized as a perturbative calculation in the original values of the field $\Psi$, 
that for the growing mode is summarized in $\delta^0$.
The integrands of the $n$-th order solution can be written in terms of the linear result $\delta^L(k,\eta)= e^{\eta-\eta_0}\delta_0(k)$ as
\be
\psi_a^{(n)}(k_1, \dots, k_n;\eta) = 
\begin{pmatrix} F_n(k_1, \dots, k_n) \\ G_n(k_1, \dots, k_n) \\ \end{pmatrix}
\delta^L (k_1,\eta) \cdots \delta^L (k_n,\eta) \, ,
\ee
where $\eta=\ln D_+(z)$ 
and the functions $F_n$ and $G_n$ fulfill the well-known recursion relations \cite{Goroff:1986ep}
\bea
\label{eq:recursionFG}
F_n(k_1, \dots , k_n) &=& \sum_{m=1}^{n-1} 
\frac{G_m(k_1,\dots,k_m)}{(2n+3)(n-1)} \times \nn \\
 && \left[ 
(2n + 1) \alpha( \kleft, \kright ) F_{n-m}(k_{m+1}, \dots, k_n) \right.  \nn \\
&& \left. \quad +  2 \beta( \kleft, \kright )  G_{n-m}(k_{m+1}, \dots, k_n)     
\right] \, ,  \\
G_n(k_1, \dots , k_n) &=& \sum_{m=1}^{n-1} 
\frac{G_m(k_1,\dots,k_m)}{(2n+3)(n-1)} \times \nn \\
 &&  \left[ 
3 \alpha( \kleft, \kright ) F_{n-m}(k_{m+1}, \dots, k_n) \right. \nn \\
&& \quad \left.+ 2n \, \beta( \kleft, \kright ) G_{n-m}(k_{m+1}, \dots, k_n)     
\right] \, ,
\eea 
where $\kleft\equiv k_1 + \dots + k_m$ and $\kright \equiv k_{m+1} + \dots + k_n$. Below we use
the $F_n$ and $G_n$ functions fully symmetrized with respect to the 
momenta and we denote them by $F_n^s(k_1,\dots , k_n)$ and $G_n^s(k_1,\dots , k_n)$.

The perturbative expansion of the power spectrum for the density field $P(k,\eta)$
can be obtained by evaluating the two functions $\Psi_{1}^{(n)}$ and $\Psi_{1}^{(n')}$ that contribute 
at order $n$ and $n'$, respectively, and summing over $n,n'\geq 1$. The corresponding contribution is
conventionally denoted by $P_{nn'}$. In general 
$P_{nn'}$ is given by a sum of terms each of which involves the product $F_{n}F_{n'}$, and explicit expressions can be obtained
by listing the different possibilities to contract the $n+n'$ Gaussian fields $\delta^L(\eta;k_i)$ contributing to $P_{nn'}$. 

In the present context we find it convenient to organize the various contributions in a slightly different way, namely with 
three indices $l, r, m$. The first two indices $l,r\geq 0$ count the contractions $\langle\delta^L(\eta;q)\delta^L(\eta;-q)\rangle$ with momentum modes belonging to one of the $F_n$ functions, respectively, and the index $m\geq 1$ counts 
the number of connections between the two $F_n$ functions. To be explicit
\bea
\label{eq:Plrm}
P_{(l,r,m)} (k,\eta) &=& \frac{(2l+m)! (2l+m)!}{2^{(l+r)} m! l! r!} e^{2(l+r+m)(\eta-\eta_0)} \int \, d\Q \, \nn \\
&& F_{2l+m}^s (k_1,\dots, k_m, q_1, -q_1, \dots, q_l, - q_l) \, \nn \\
&& F_{2r+m}^s (-k_1,\dots, -k_m, p_1, -p_1, \dots, p_r, - p_r) \, , \, 
\eea
where the integration measure is given by 
\bea
\label{eq:measure}
\int d\Q &=& \left( \prod_i \int  \, d^3\Q_i \, P^0(\Q_i) \right) \, \delta^{(3)}(k - \sum_{i=1}^m k_i) \, .
\eea
Here $P^0(k)$ denotes the initial power spectrum at $\eta=\eta_0$
and the $\Q_i$ run over $k_1 \dots k_m$, $q_1 \dots q_l$ and $p_1 \dots p_r$. The full power spectrum
is given by the sum over all terms with $l,r\geq 0$ and $m\geq 1$. We note that the conventional $P_{nn'}$
are given by summing over all the $P_{(l,r,m)}$ with $n=2l+m$ and $n'=2r+m$. Each contribution $P_{(l,r,m)}$ can be
interpreted as a Feynman diagram featuring two kernels denoting the $F_n$-functions, that are connected by $m$ lines,
and have $l$ and $r$ lines that are starting and ending at the same kernel function, respectively.
Consequently, $P_{(l,r,m)}$ contributes to the power spectrum at the $L=l+r+m-1$ loop order.

The tree-level and one-, two- and
three-loop contributions are in this notation given by
\bea
P_{lin} &=& P_{(0,0,1)} = e^{2(\eta-\eta_0)}P^0(k) \, , \nn \\  
P_{1-loop} &=& P_{(0,0,2)} + 2 P_{(0,1,1)} \, , \nn \\  
P_{2-loop} &=& P_{(0,0,3)} + 2 P_{(0,1,2)} + P_{(1,1,1)} + 2 P_{(0,2,1)} \, , \nn \\  
P_{3-loop} &=& P_{(0,0,4)} + 2 P_{(0,1,3)} + P_{(1,1,2)} + 2 P_{(0,2,2)} \nn \\ 
&&  + 2 P_{(1,2,1)} +  2 P_{(0,3,1)} \, .  
\eea
Compared to the standard notation, $P_{22}=P_{(0,0,2)}$, $P_{13}=P_{(0,1,1)}$,
$P_{24}=P_{(0,1,2)}$, $P_{15}=P_{(0,2,1)}$, $P_{33}=P_{(0,0,3)}+P_{(1,1,1)}$, etc.

\subsection{Efficient evaluation at any loop}

An efficient evaluation of the power spectrum is typically hindered by two factors. First, 
there is a cancellation between different big contributions in the limit of large external momentum $k$.
This cancellation has to be dealt with at the level of the integrand for a reasonable accuracy. Second, 
the integrand contains a large number of terms and performance is an issue in the numerical 
evaluation of the integrand. 

The first problem can be overcome with the procedure presented in \cite{Blas:2013bpa} and also discussed in detail in 
the context of the effective theory approach to cosmological perturbation in \cite{Carrasco:2013sva}. In the following, 
we briefly review this approach. The cancellation under consideration stems from contributions that arise if some of the 
loop momenta $\Q_i$ in Eq.~\eqref{eq:measure} become soft. In this regime the functions $F_n^s$ are enhanced
by a factor $\propto k/|Q_i|$ for each soft momentum. However, after summing 
over all contributions and after integration the enhancement is absent. This absence follows from Galilean invariance
\cite{Scoccimarro:1995if, Kehagias:2013yd, Peloso:2013zw}
and was proven in detail in \cite{Blas:2013bpa}. 

Remarkably, the cancellation can be made
explicit already at the level of the integrand.
To understand how to do it, let us start with the  one-loop example, where one has two contributions of the form
\bea
2P_{(1,0,1)} &\propto& 2 \int d^3 k_1 d^3q_1 F^s_3(k_1, q_1, - q_1) F^s_1(-k_1) P^0(q_1) P^0(k_1) \delta^{(3)}(k_1 - k)  \nn \\
&=& 2 \int d^3 q_1 F^s_3(k, q_1, - q_1)  P^0(q_1) P^0(k) \, , \nn 
\eea
and
\bea
\label{P002}
P_{(0,0,2)} &\propto& \int d^3 k_1 d^3k_2 [ F^s_2(k_1, k_2)]^2 P^0(k_1) P^0(k_2) \delta^{(3)}(k_1 + k_2 - k) \nn \\
 &=& \int d^3k_2 [F^s_2(k - k_2, k_2)]^2 P^0(k - k_2) P^0(k_2) \, .  
\eea
The first term experiences an enhancement for soft internal momentum $q_1$, while the second term is enhanced for 
either $k_2$ soft or $(k-k_2)$ soft. Even though the final result has no enhancement by soft modes, 
different regions in the integration conspire to cancel each other. This problem can be avoided 
by enforcing that $(k-k_2)$ cannot become soft. Since the integrand is symmetric under 
$(k - k_2) \leftrightarrow k_2$ (inherited from $k_1 \leftrightarrow k_2$) this can be achieved by 
inserting a factor $\Theta(|k-k_2| - |k_2|)$ (respectively $\Theta(|k_1| - |k_2|)$) in Eq.~\eqref{P002} 
and compensating by a factor $2$. After identifying the one loop momentum $\Q_1 \equiv q_1 \equiv k_2$ 
and symmetrizing $\Q_1 \leftrightarrow -\Q_1$ the integrand is not enhanced in the soft regime if 
both contributions are added (the partial enhancement arises from the same region of integration).

This procedure is readily generalized to higher orders. For any $P_{(l,r,m)}$ with $m\geq 2$ one can use the symmetry in
the momenta $k_1,\dots,k_m$ to single out $k_1$ to be the largest 
loop momentum and remove it via integration over the delta function after inserting a factor
\be
\label{T1}
\left. m \, \prod_{i=2}^m \Theta(|k_1| - |k_i|) \right|_{k_1 = k - k_2 - \dots - k_m} \, .
\ee
Alternatively, one can also sort all the loop momenta
\be
\label{T2}
\left. m! \, \prod_{i=2}^{m} \Theta(|k_{i-1}| - |k_{i}|) \right|_{k_1 = k - k_2 - \dots - k_m} \, ,
\ee
what is obviously equivalent since the integrand is fully symmetric in the momenta $k_i$.
In both cases, the integrand then depends on the external momentum $k$ as well as on the
loop momenta $\Q_i = k_2,\dots,k_m,q_1,\dots,q_l,p_1,\dots,p_r$. Here the index $i$ can be chosen
to run from $1$ to $L=l+r+m-1$, the number of loop momenta.  The possible enhancement
will now only follow from soft  modes in the loop momenta $\Q_i$. In the next step,
the integrand should be symmetrized with respect to arbitrary permutations of the $\Q_i$, to ensure that
all the internal momenta are treated on the same footing.
Similarly, one has to symmetrize the integrand with respect to the sign-flips $\Q_i \leftrightarrow - \Q_i$ of any of the loop momenta.
After these manipulations, the resulting integrands for all $P_{(l,r,m)}$ with indices satisfying $L=l+r+m-1$ should be added to obtain an expression for the $L$-loop integrand. 

Both choices \eqref{T1} and \eqref{T2} lead to an infrared safe integrand. 
We tested both and find that the second is slightly more stable in the numerical integration due to a less redundant 
integration region. As in the one-loop case, we observed that the integrand is not enhanced for soft modes 
if all contributions at fixed loop order are summed over. We tested this analytically up to two loops and 
numerically up to four loops as already reported in \cite{Blas:2013bpa}.

In conclusion, the expression at $L$-loop evaluated in the numerics is 
\be
P_{L-loop} = e^{2(L+1)(\eta-\eta_0)} \int \, d^3\Q_1 
\cdots d^3\Q_{L} P^0(\Q_1)\dots P^0(\Q_L) \sum_{
\genfrac{}{}{0 pt}{}{l,r\geq0,m\geq 1}{m+l+r = L + 1}
} I_{(l,r,m)} 
\ee
with the integrand
\bea
\label{eq:integrand}
I_{(l,r,m)} &=& \frac{(2l+m)! (2r+m)!}{2^{(l+r)} l! r! m!} \
 {\rm Symm} \Bigg[\bigg\{ \ m!\, \prod_{i=2}^{m} \Theta(|k_{i-1}| - |k_{i}|) \times \nn \\
&&\hspace{-.8cm}  P^0(k_1)F_{2l+m}^s (k_1,\dots, k_m, q_1, -q_1, \dots, q_l, - q_l) \times  \, \nn \\
&&\hspace{-.8cm}   F_{2r+m}^s (-k_1,\dots, -k_m, p_1, -p_1, \dots, p_r, - p_r)  \, 
\bigg\}_{k_1 = k - \sum_{j=2}^m k_j} \Bigg] \, .
\eea
The symmetrization denotes a sum over all $N_{(l,r,m)}\equiv\frac{L!}{(m-1)! l! r!}\times 2^{m-1}$ possibilities to choose the momenta 
$k_2 \dots k_m$, $q_1 \dots q_l$ and $p_1 \dots p_r$ out of the $L$ momenta $\Q_1 \dots \Q_L$ as well as performing the sign changes 
$k_j \leftrightarrow - k_j$ for $j=2,\dots,m$, multiplied by a normalization factor $1/N_{(l,r,m)}$. Note that the symmetrizations affect also the
first argument of the $F_n$ functions since $k_1\equiv k - \sum_{j=2}^m k_j$. 

Unfortunately, this procedure tremendously increases the number of terms one has to evaluate.
In particular, the symmetrization of the integrand in (\ref{eq:integrand}) generates a large number of contributions. It is thus essential to only perform the necessary symmetrizations. In Eq.~\eqref{eq:integrand} we 
already used the fact that $F^s_n$  is symmetric in its parameters. This implies that 
one should only average over the 
$\frac{L!}{(m-1)! l! r!}$ terms arising from picking the $k$, $q$ and $p$ momenta out of the set of all $\Q$ (instead of all $L!$ orderings). 
Likewise, the symmetric $F^s_n$ and $G^s_n$ can be determined from (\ref{eq:recursionFG}) by summing over 
$\binom{n}{m}$ terms if the functions $F_n$ and $G_n$ on the right-hand side are already the symmetric ones.

Since the final evaluation of $F_n$ and $G_n$ only depends on the scalar product between different vectors, 
it is very efficient to pre-calculate and store partial results on different stages. The vectors that appear in the scalar products are
of the form
\be
\label{eq:vec_linear}
v = c_k \, k + \sum c_i \, \Q_i \, ,
\ee
where the $c$ can be $\pm 1$ or $0$. By convention one can also chose $c_k$ to be non-negative. At $L$-loop, there are
$n_L=2\cdot 3^L$ vectors of this type, e.g. $n_3=54$ at three loop order. It is much faster to enumerate the existing
linear combinations in the beginning and to store the pre-calculated values of the $n_L(n_L+1)/2$ scalar products in a table
rather than using the real-valued vectors in the recursion. Since all vectors that appear in the evaluation of the integrand
are of the form (\ref{eq:vec_linear}), they can be represented by a vector of length $L+1$ of the coefficients $(c_k,c_1,\dots, c_L)$ where the $c_i$  are elements of $Z_3$. Adding and subtracting vectors within this class can then be handled efficiently via the basic operations modulo 3.

At the same token, also the required $F^s_n$ and $G^s_n$ functions can be pre-calculated. Their arguments are the loop momenta $\Q_i$
(with a sign) and a vector of the kind as (\ref{eq:vec_linear}) with $c_k=1$ (cf. \eqref{eq:integrand}). Since the functions $F^s_n$ and $G^s_n$ are 
symmetric under interchanging the momenta, only $(3^L+1) 4^L$  different combinations can appear as 
parameters. The factor $(3^L+1)$ stores the form of the vector involving $k$ as in (\ref{eq:vec_linear}) 
[since $c_k=1$ there are only $3^L$ vectors that appear, plus one possibility that no vector of this form is present in the argument]; 
the factor $4^L=2^L\cdot 2^L$ stores the information if the loop momentum $\Q_i$ is present or not as an argument ($2^L$ possibilities), and similarly whether $-\Q_i$ is present (also $2^L$ possibilities). Storing the $F^s_n$ 
and $G^s_n$ functions with the required sets of arguments in a table improves the performance tremendously, especially  because the functions with a low number or arguments (up to $\sim 5$ at three loops; the maximum is $7$) are typically evaluated many times due to the recursive calculation.

In the presented data, the Monte Carlo integration library CUBA \cite{Hahn:2004fe} has been used. 
We evaluate the eight-dimensional 
integrals with up to $10^8$ evaluations of the integrand. The errors shown are the ones resulting from the 
numerical integrations. We performed numerous checks using two completely independent codes. In particular, we tested 
different parametrizations of the loop momenta. 
Some more details about the numerical integration can be found in App.~\ref{app:numerics}.

\section{Three-loop results\label{sec:3Lresults}}

\subsection{Expectations}

Before presenting our numerical results and  implications, we would like 
to discuss the expectations in the large $k$ and small $k$ regimes as analyzed in ref.~\cite{Blas:2013bpa}. 
In both regimes the variance of the density field \cite{Scoccimarro:1995if}
\be
\label{eq:NLFS}
\sigma^2_l(k,z)\equiv 4\pi \int_0^k dq\,q^2 P_{lin}(q,z) = 4\pi D_+(z)^2 \int_0^k dq\,q^2 P_{lin}(q,z=0),
\ee
plays an important role. In the previous expression $P_{lin}$ is the power spectrum at linear level.  At large $k$ and for an initial power spectrum similar to Eisenstein-Hu \cite{Eisenstein:1997jh}, the leading
logarithmic behavior in $k$ is given at $L$-loop order by contributions of the form ($n \leq 2L$)
\be\label{eq:largek}
P_{L-loop} \ni \left([ k \partial_k ]^n P_{lin}(k,z) \right)\,\sigma_l^{2L}(k,z) \, .
\ee
Subleading logarithms can give sizable corrections \cite{Blas:2013bpa}. 
For small $k$ on the other hand, one finds
\be
\label{eq:smallk}
P_{L-loop} \to  - \frac{61}{105}\, C_L\, k^2 P_{lin}(k,z)\, \frac{4\pi}{3} \int_0^\infty dq P_{lin}(q,z) \,\sigma_l^{2L-2}(q,z)\,.
\ee
The convergence of the expansion in loops depends ultimately on the quantity $\sigma_l(q,z)$ and on the coefficients $C_L$ that are 
unknown (the normalization has been chosen such that at one-loop $C_1=1$). For the first three loop orders they
are order one.

The problem is that $\sigma_l$ is sensitive to the UV part of the power spectrum what hinders 
the convergence of SPT also for soft momenta $k$. Parametrically, it scales for a Eisenstein-Hu spectrum as
\be
\sigma_l^2(k,z) \simeq  D_+(z)^2\int^k \frac{d^3q}{q^3} \log^2(e + q/k_0)\, \simeq
 D_+(z)^2\log^3(e + k/k_0)\,   \, .  
\ee
Due to the logarithmic growth, the $q$-integral in Eq.~(\ref{eq:smallk}) is a convergent integral for any $L$ (note that $P_{lin} \to q^{-3}\log^2(e + q/k_0)$). Therefore, each of the loop integrals is finite, and there is no need to introduce a UV cutoff\footnote{On practical grounds, it is necessary to introduce a cutoff for numerical calculations, which we chose large enough to capture the complete integral (see App. \ref{app:numerics}).}.
Consequently, for small $k$ the $L$-loop contributions scale like
\bea\label{eq:smallkScaling}
C_L\,\int_0^\infty dq P_{lin}(q,z) \,\sigma_l^{2L-2}(q,z) 
&\simeq& C_L D_+(z)^{2L}\int dl \, e^{-2 l} l^{3L-1} \,   \nn\\
&\simeq&\frac{(3L-1)!}{ 2^{3L}} \, C_L D_+(z)^{2L}\, .
\eea
The first factor in the second line of Eq.~\eqref{eq:smallkScaling} grows very fast 
with the number of loops. So even for large redshift (for which there is an additional suppression $D_+\sim1/(1+z)$), the convergence of the loop series can be at best asymptotic\footnote{Another famous example of a perturbative series that is strictly non-convergent, but asymptotically converging is the the loop expansion in QED \cite{Dyson:1952tj}.}  unless the coefficients $C_L$ in 
(\ref{eq:smallk}) do produce a strong suppression (we do not find an indication for such a suppression up to three loop
order).

\subsection{Numerical Results and Implications}
\begin{figure}[t]
\begin{center}
\includegraphics[width=0.75\textwidth]{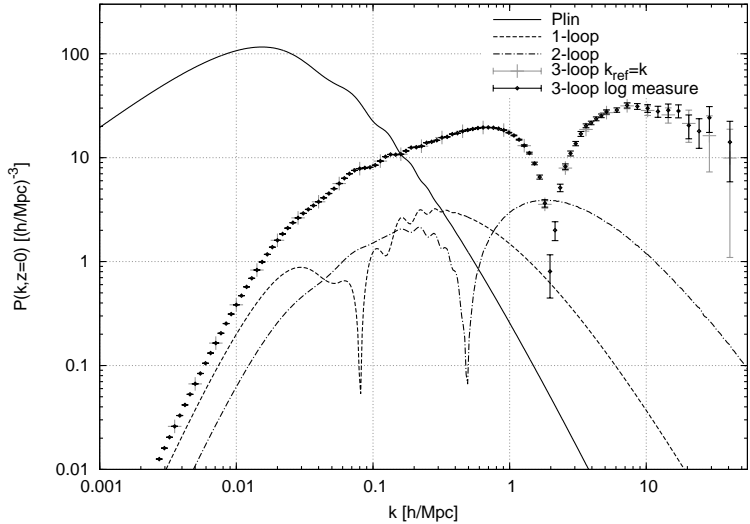}
\end{center}
\caption{\label{fig:power3L} One, two and three-loop contributions to the equal-time power spectrum obtained from
a numerical Monte Carlo integration within standard perturbation theory at $z=0$. The linear power spectrum is obtained
from the initial power spectrum from CAMB \cite{Lewis:1999bs} using the $\Lambda$CDM model with WMAP5 parameters. For the three-loop order, the error bars show an estimate
for the numerical error obtained by multiplying the error output of the CUBA routine Suave by a factor of two. The relative error is $\leq 0.002$ for $k\leq 0.55\, h/$Mpc. The black diamonds and grey crosses correspond to two different parametrizations of the absolute loop momenta (see App.~\ref{app:numerics}).}
\end{figure}

Figure \ref{fig:power3L} shows the power spectrum up to three loops at $z=0$. 
One observes that even for very small $k$ the three-loop result is larger than even the one-loop
term. This indicates that SPT does not converge in this regime even though 
the linear contribution dominates over the subleading ones for $k \to 0$. This was already
observed in \cite{Bernardeau:2012ux} where the propagator in SPT was studied at three-loop 
order. Given an asymptotic series, its form may still provide very relevant
information about the non-linear behavior of the solution \cite{Bender:Book}. 
We comment on a possible way of achieving this through a resummation of the different contributions below.

Another observation is that for $z=0$ the sum of loop corrections up to three loops becomes larger than
the linear power spectrum for $k \gtrsim 0.16 \, h/$Mpc. Since the former is negative, SPT clearly does
not converge neither on these scales. For even larger momentum $k$, one observes that each loop contribution
features the expected behavior (\ref{eq:largek}) with a logarithmic enhancement compared to the
linear spectrum. But also in this regime, the loop expansion appears to be divergent.

\begin{figure}[t]
\begin{center}
\includegraphics[width=0.425\textwidth]{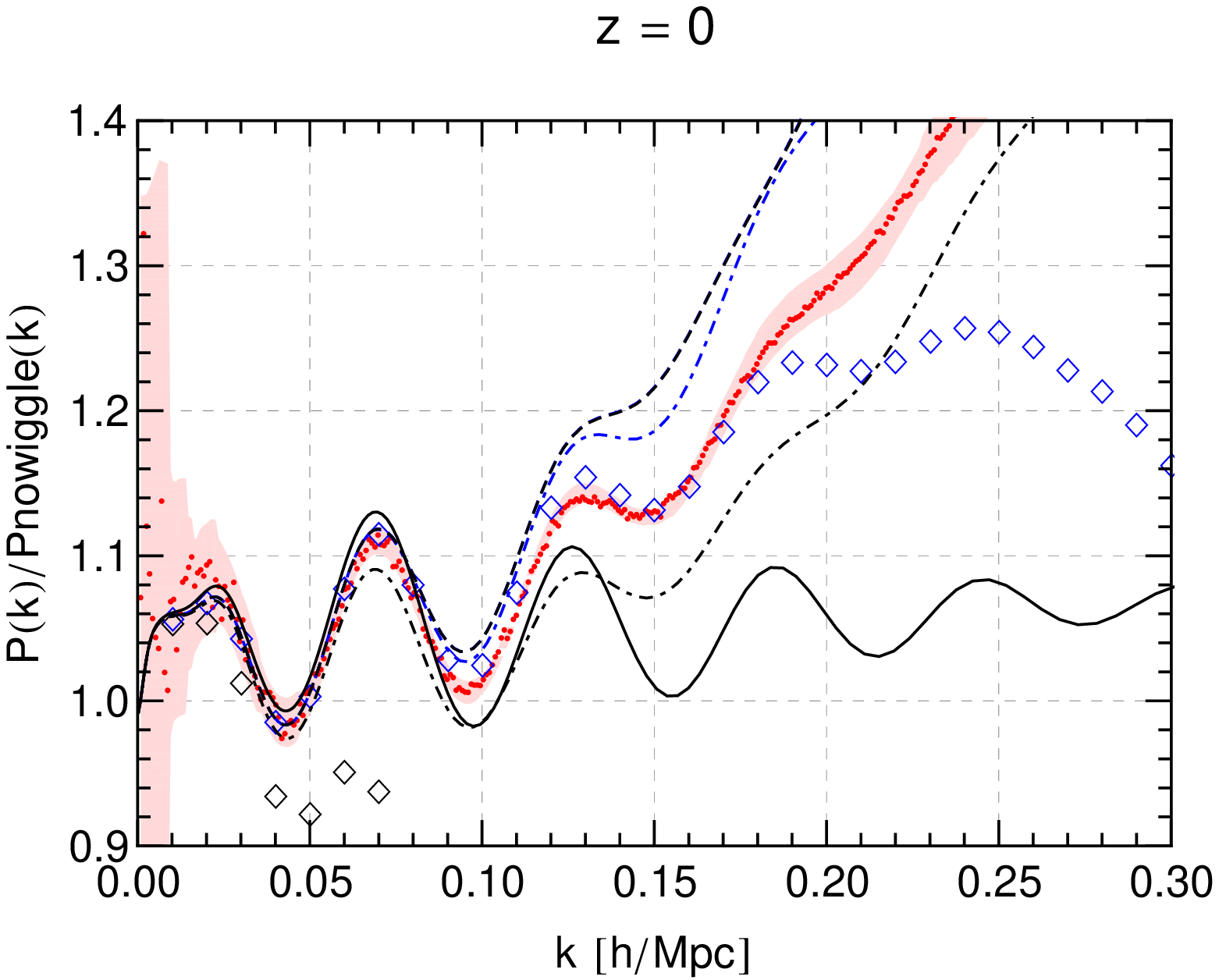}
\includegraphics[width=0.425\textwidth]{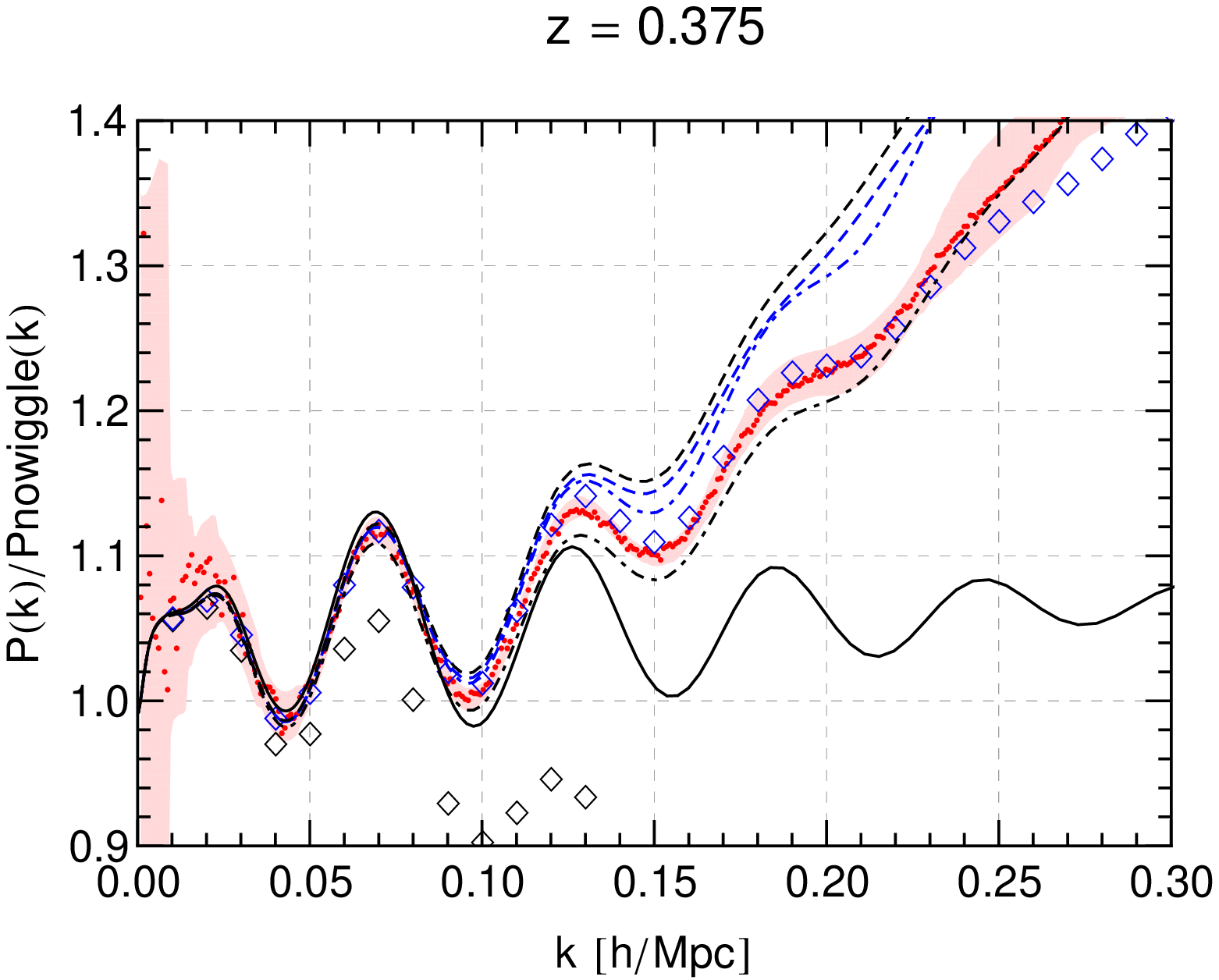}
\\
\includegraphics[width=0.425\textwidth]{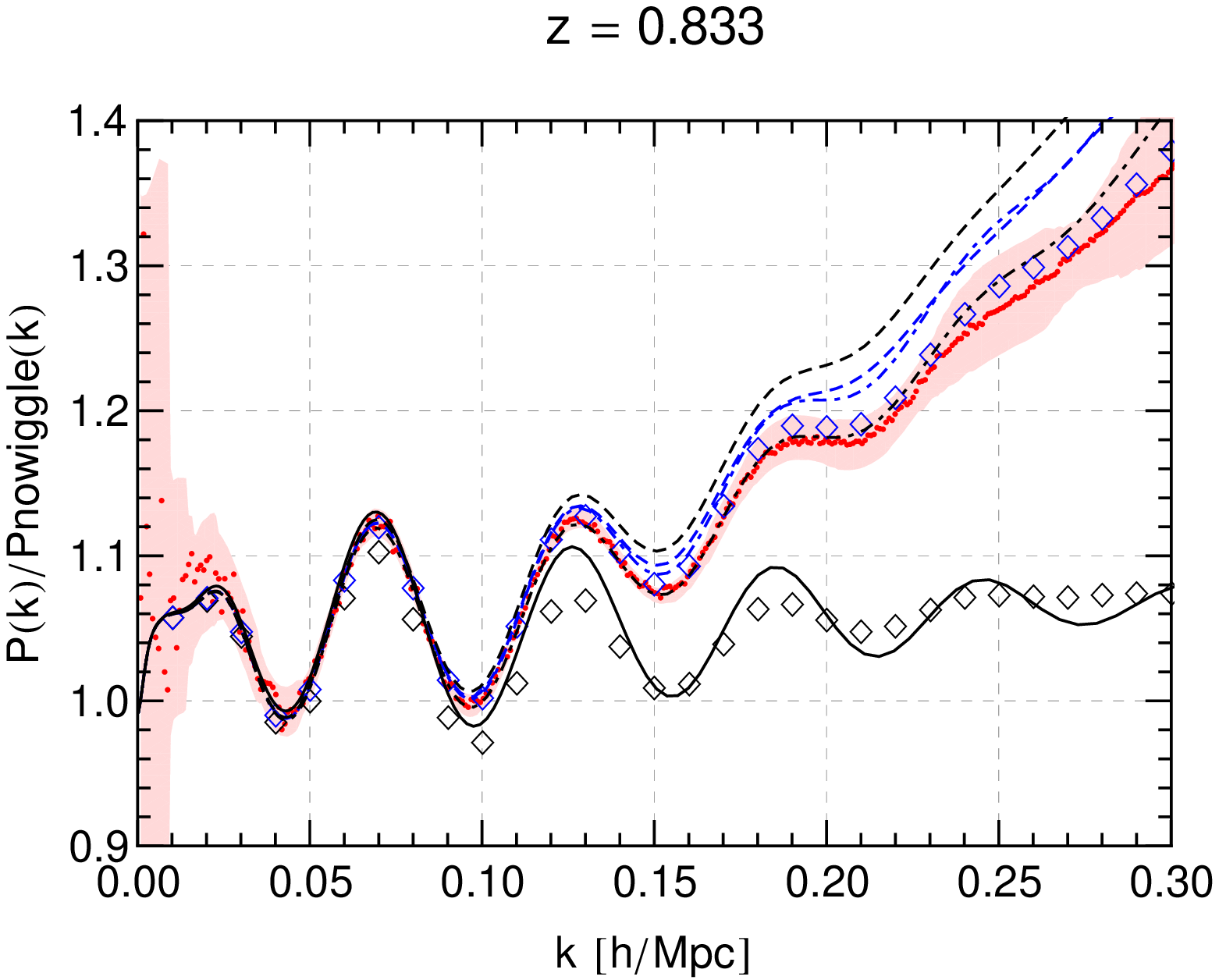}
\includegraphics[width=0.425\textwidth]{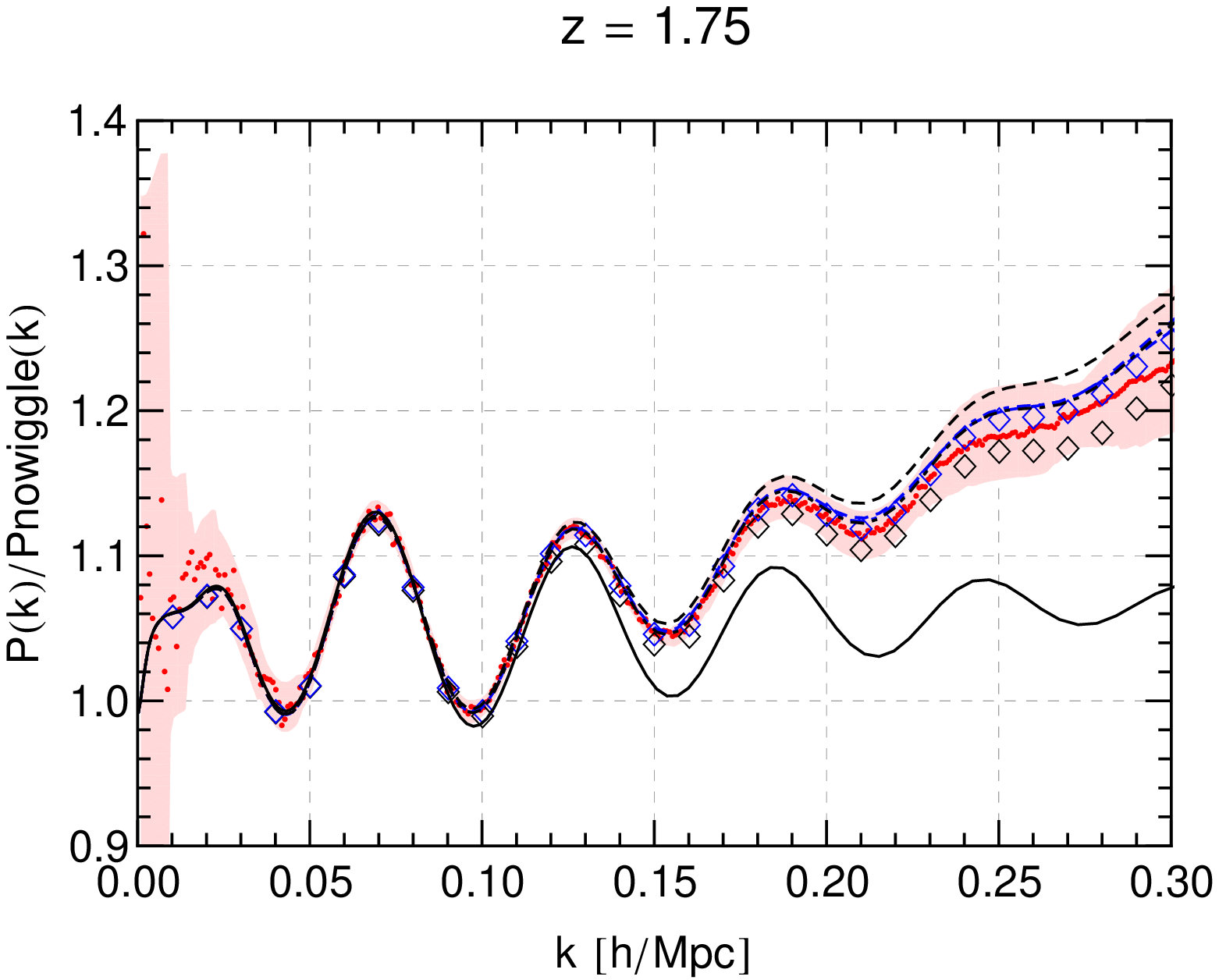}
\end{center}
\caption{\label{fig:nbody_lowz} Comparison at redshifts $z = \{0, 0.375, 0.833, 1.75\}$ of SPT up to one 
loop (black dashed lines), two loops (black dot-dashed) and three loops 
(black diamonds) with N-body results of the Horizon Run 2~\cite{Kim:2011ab} (red dots, see App.~\ref{app:hr3}). The black line
corresponds to the linear result. We also show the results 
of  \pade resummation (same styles as for SPT but in blue, see Sec.~\ref{sec:Pade});
at $z=0$ the blue and black dashed line lie on top of each other.}
\end{figure}

\begin{figure}[t]
\begin{center}
\includegraphics[width=0.425\textwidth]{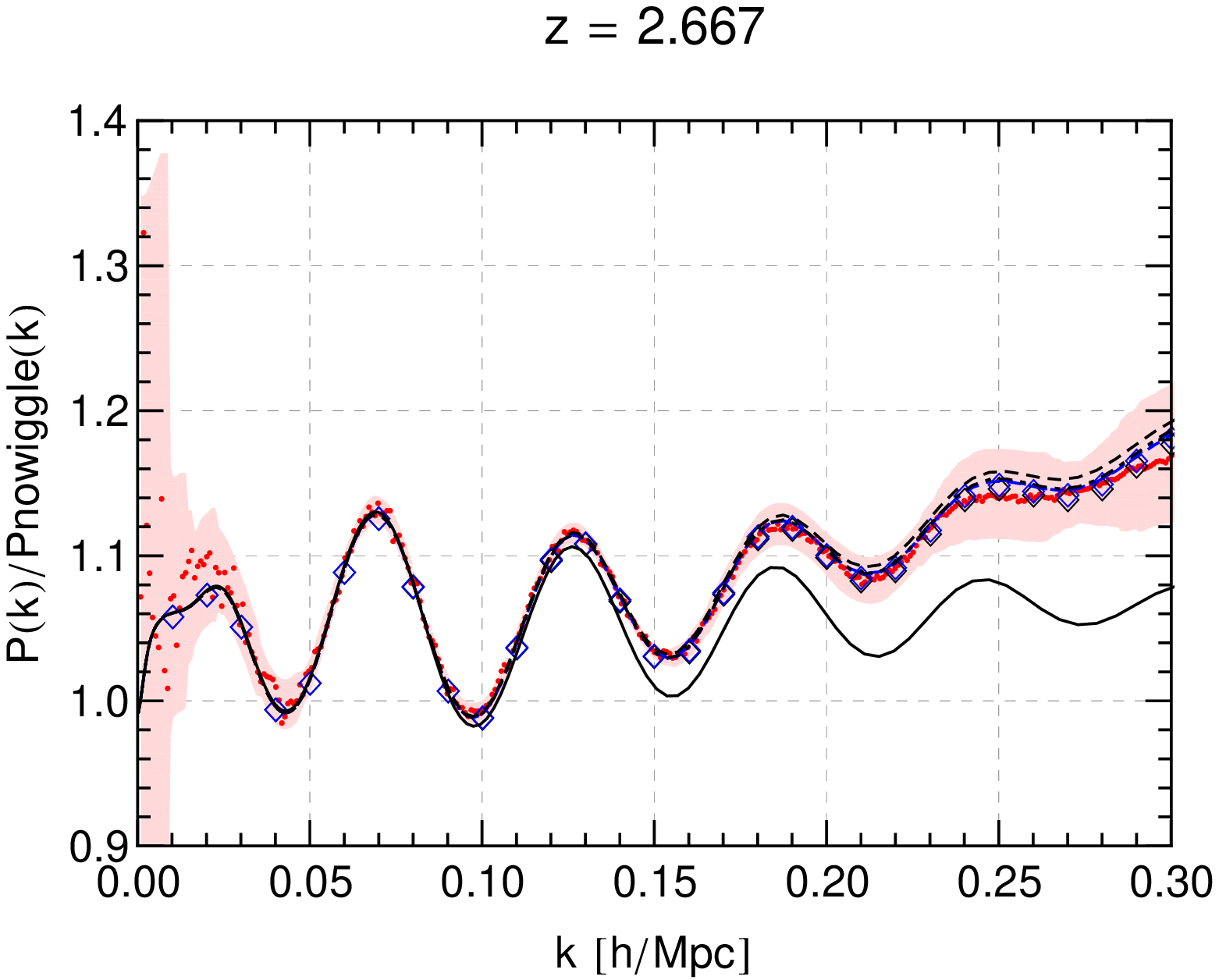}
\includegraphics[width=0.425\textwidth]{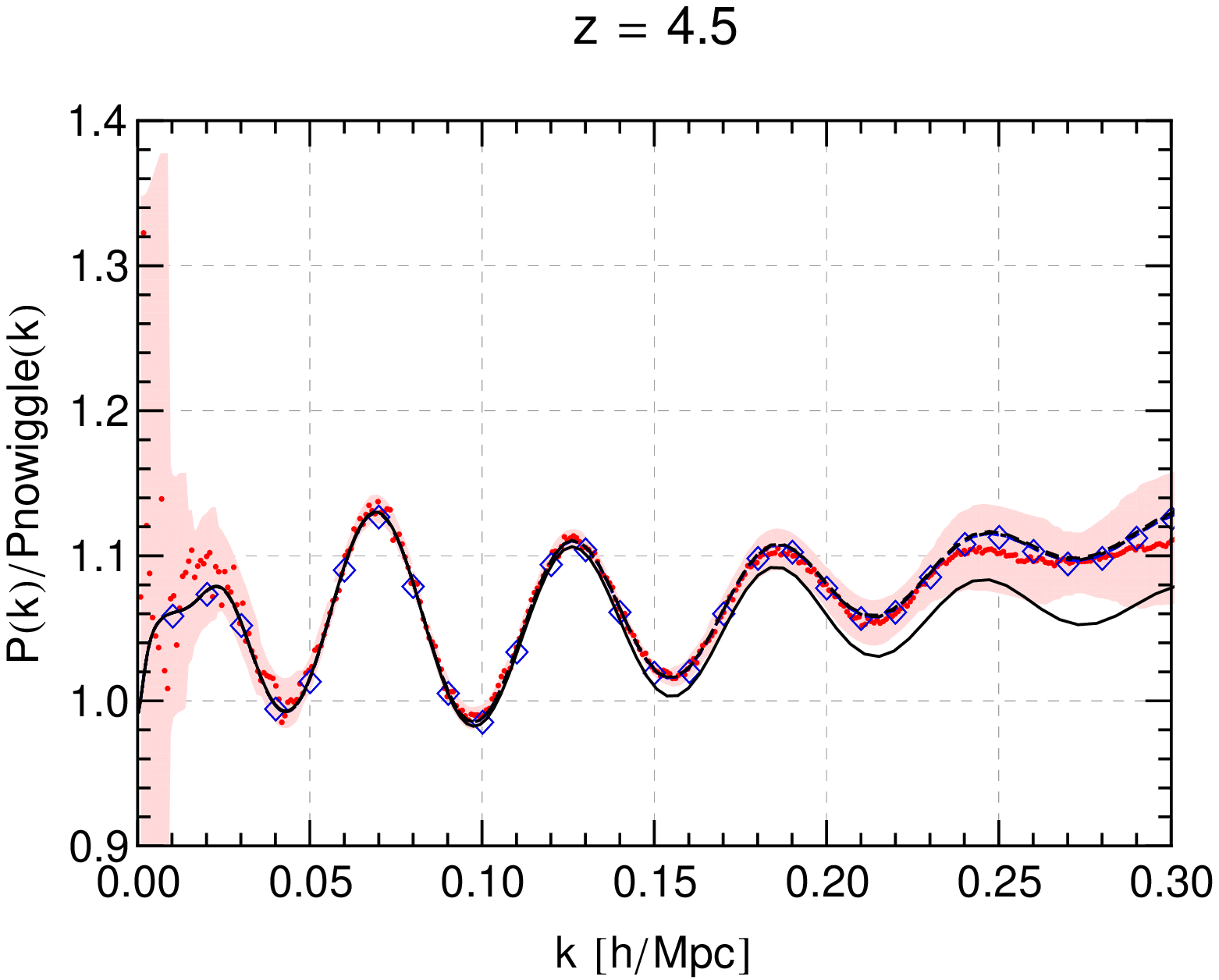}
\end{center}
\caption{\label{fig:nbody_highz} Same as Fig.~\ref{fig:nbody_lowz} for redshifts  $z = \{2.67, 4.5 \}$ }
\end{figure}

The picture might change if one goes to larger redshift $z$, where the expansion parameter can be efficiently suppressed since 
$\sigma^2_l \propto D_+(z)^2 \sim (1+z)^{-2}$. In Figs.~\ref{fig:nbody_lowz} and \ref{fig:nbody_highz} we show some comparisons between our three-loop
SPT results (black lines and diamonds) and N-body simulations (red dots, Horizon Run 2 \cite{Kim:2011ab}) for various redshifts (see App.~\ref{app:hr3} for further details). For large redshift ($z\gtrsim 1.75$) the three-loop contribution may lead to an improved agreement with the N-body data, while it clearly degrades the agreement compared to the two-loop at lower redshifts. The same happens for the two-loop at even smaller redshifts and at small momenta. This indicates that for any redshift, adding loop contributions improves the agreement only up to a certain order, as typically expected for asymptotically converging series.

\begin{figure}[t]
\begin{center}
\includegraphics[width=0.75\textwidth]{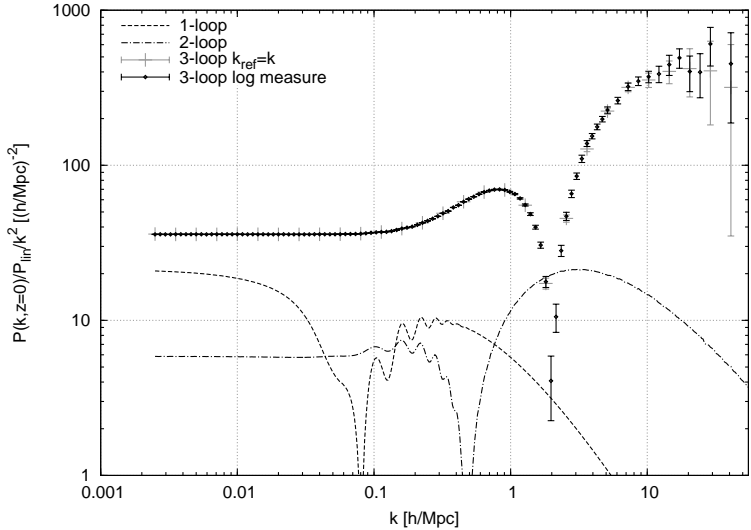}
\end{center}
\caption{\label{fig:power3LNorm} Ratio $P_{L-loop}(k,z=0)/P_{lin}(k,z=0)/k^2$ for the one- two- and three-loop contributions (line styles as in Fig.~\ref{fig:power3L}).}
\end{figure}

In general, in such a situation, one expects that the partial sum up to the smallest term yields the most accurate estimate of the full
result, with a theoretical uncertainty of the order of the smallest term. For a realistic initial power spectrum,
this indicates that the power spectrum at $z\lesssim 1$ can be estimated with SPT at most to an accuracy of the order of the two-loop contribution (e.g. $P_{2-loop}/P_{lin}\simeq 6\%$ at $z=0$ and $k=0.1\, h/$Mpc).

As already emphasized, this does not mean that it is  in principle impossible to achieve a better precision within this framework. Indeed, there are well-known
examples where a resummation of an asymptotically diverging perturbative series yields physically meaningful results, provided the leading behavior of the higher orders is known or can be estimated, see e.g.~\cite{Dunne:2004nc,Moore:2000jw}. For the problem
of gravitational collapse, some approaches already exist for the case of non-linear evolution of collapsing spheroidal bodies \cite{Tatekawa:2006gx,Matsubara:1997bx} and in the context of Lagrangian perturbation theory (LPT) \cite{NadkarniGhosh:2010th}.

\bigskip

It is interesting to compare the SPT dynamics studied here with other approaches to the problem
of non-linear behavior of cosmological perturbations. Some of these approaches reproduce the low-$k$ limit of SPT (e.g. 
\cite{Anselmi:2012cn,Taruya:2012ut}), and the same asymptotic behavior at low-$k$ is expected. Accordingly, these approaches cannot converge in the regime of low momenta. This is not obvious for other approaches, as the Zel'dovich approximation (ZA). For the latter, the structure  of the $F_n$ kernels differs from the SPT dynamics at low-$k$, and different contributions at higher loop order are suppressed by additional factors of the momentum $k$. This is clearly seen in Fig.~1 of \cite{Crocce:2005xy}, where the calculation is performed to three-loops\footnote{We have also tested our numerical code against this result.}. One may wonder if this approach neglects important contributions or if it even 
properly resums the asymptotic behavior of SPT. A possible way to study this is by developing a perturbation theory around ZA. This can be readily done in LPT. In this case, the results up to two-loops can be found in \cite{Okamura:2011nu}. Unfortunately, the individual loop contributions to the power spectrum are not presented such that the convergence properties cannot be easily inferred (For a comparisons of ZA and
the asymptotic behavior of SPT dynamics in the context of spherical collapse see \cite{Matsubara:1997bx}). 

\section{\pade resummation~\label{sec:Pade}}

In the previous section, we discussed strong indications that SPT does not provide a convergent expansion for the power spectrum at low-$k$. In fact, the inclusion of three-loop results makes the expansion look asymptotic, with the one-loop result
representing the optimal fit to the data at $z=0$. We also noticed that the same formalism fails at any order to give reliable predictions for 
the behavior of the power spectrum at scales where the baryonic acoustic oscillations (BAO) are today (cf. Fig.~\ref{fig:nbody_lowz}) \cite{Percival:2007yw}. In this section we want to explore the possibility that these problems are related. Namely, we use a \pade ansatz to resum the low-$k$ result and see if it can 
provide a perturbation theory that leads to reliable predictions in the BAO regime.

\subsection{\pade resummation for the low momentum kernels}

Our starting point is the observation
that the range of $k$ values for which the small-$k$ limit (\ref{eq:smallk}) is valid increases with increasing loop order. This can be seen clearly in Fig.~\ref{fig:power3LNorm}, where we show the loop contributions to the power spectrum normalized to $k^2 P_{lin}(k,z)$. In particular, this reveals that $P_{3-loop}\propto k^2 P_{lin}(k,z)$ up to $k\lesssim 0.08\, h$/Mpc to percent accuracy, while the two-loop starts to deviate from this limit by more than one percent already at $k\sim 0.06\,h$/Mpc and the one-loop at $k\sim 0.003\,h$/Mpc. 
Supposing that also the higher loop orders can be well-described by the `small-$k$' limit up to similar momenta as the two- and three-loop, this motivates to investigate the divergent loop series in this limiting regime more closely. 

\begin{figure}[t]
\begin{center}
\includegraphics[width=0.49\textwidth]{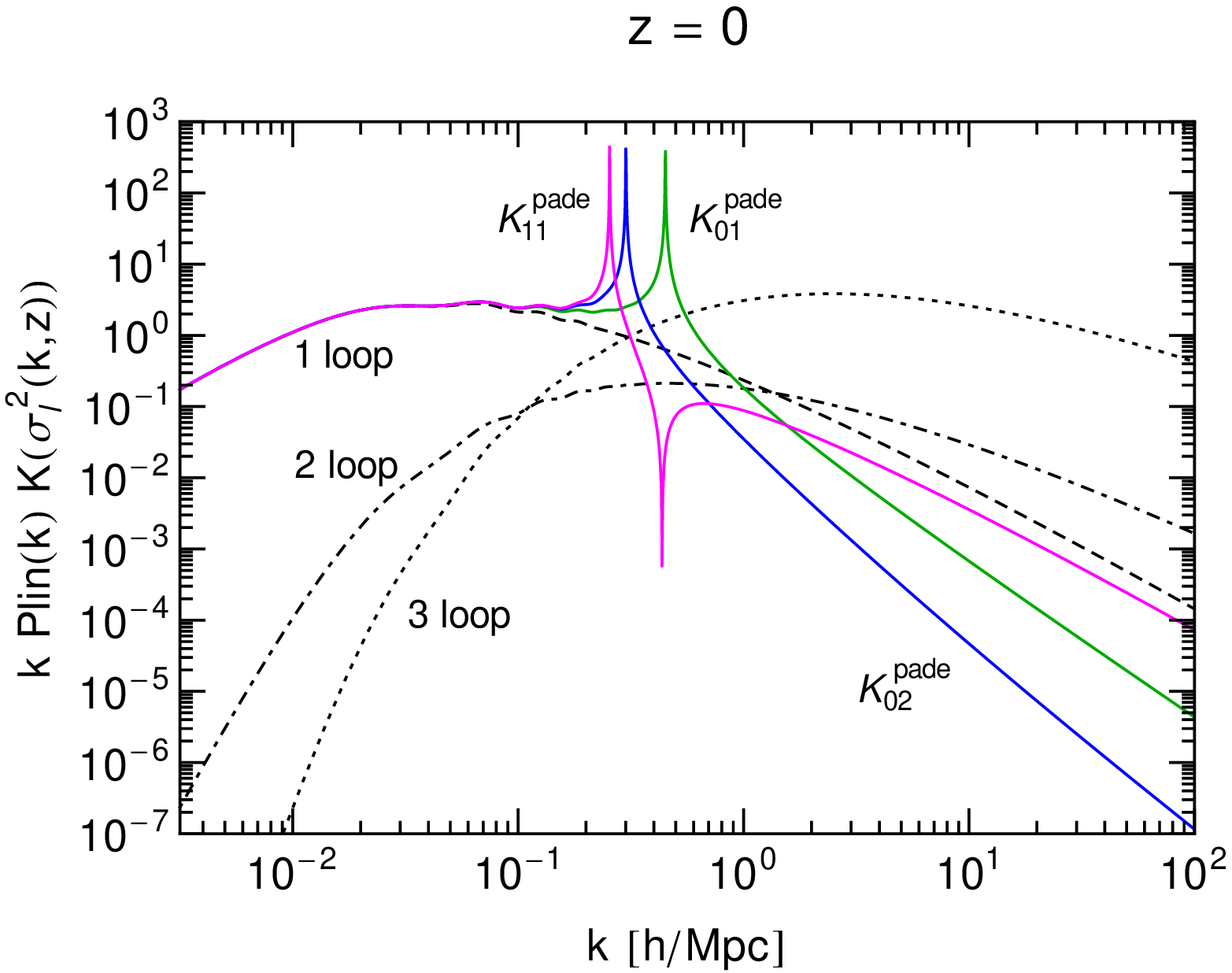}
\includegraphics[width=0.49\textwidth]{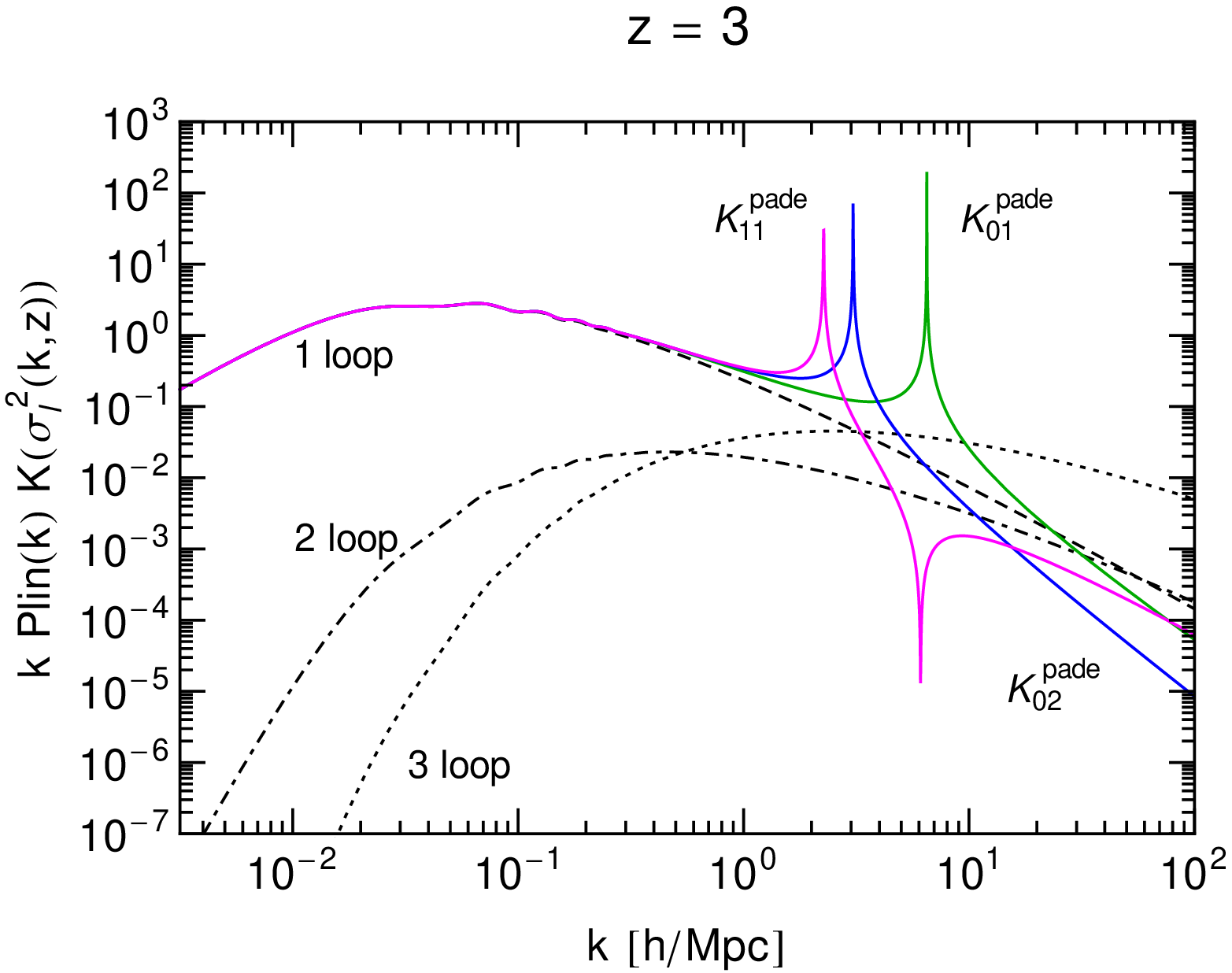}
\end{center}
\caption{\label{fig:padeKernel} Integrand kernel $k\, P_{lin}(k) K_L(\sigma_l^2(k,z))$ for the power spectrum as obtained in SPT at
one-loop (black dashed), two loops (black dot-dashed), three loops (black dotted). The solid lines are the integrand kernels
obtained after \pade resummation, $K^{pade}_{01}$ (green), $K^{pade}_{02}$ (blue) and $K^{pade}_{11}$ (magenta). The factor $k\, P_{lin}(k)$
is chosen such that the area under the curves represents the integral when using a logarithmic integration measure.}
\end{figure}

For the linear power spectrum corresponding to the WMAP5 parameters used in \cite{Kim:2011ab}, we find that
the coefficients in the small-$k$ expansion (\ref{eq:smallk}) up to three-loop order are given by  $C_1=1$, $C_2\simeq 0.71$, and $C_3\simeq 1.05$
(see App.~\ref{app:pade} for analytic expressions at $L$-loop).
The full result for the power spectrum in the small-$k$ limit can be written as
\be\label{eq:Psmallk}
P_{small-k}(k,z) \equiv - \frac{61}{105} k^2 P_{lin}(k,z) \frac{4\pi}{3} \int_0^\infty dq P_{lin}(q,z) \,K(\sigma_l^{2}(q,z)),
\ee
where the integrand kernel is given by a series in $x\equiv \sigma_l^2(q,z)$, 
\be\label{eq:kloop}
K(x) = \sum_{L=1}^\infty C_L x^{L-1} \;.
\ee
The divergent behavior of the loop expansion originates from the increasing powers of $x$ inside
the $q$-integral in (\ref{eq:smallk}), as  discussed before. Therefore, a resummation of this series could remedy the
divergence in the small-$k$ limit. However, this would require some knowledge of the
asymptotic behavior of the $C_L$. In the following, we will explore the consequences of using a \pade ansatz (see e.g. 
\cite{Bender:Book} for a discussion of \pade resummations\footnote{Other examples
of the use of \pade ansatz in physics can be found in \cite{Dunne:2004nc,Moore:2000jw,Pade:Book}. See also \cite{Matsubara:1997bx}
for results in LPT.}) of the form
\be
K_{nm}^{pade}(x) \equiv \frac{1+\sum_{i=1}^n a_i x^i}{1+\sum_{j=1}^m b_j x^j} \,,
\ee
which satisfies the normalization condition $K(0)=1$. The coefficients $a_i$ and $b_j$ can be
determined by matching the Taylor coefficients of the \pade ansatz to the perturbative SPT calculation.
When taking only the one- and two-loop coefficients into account, one obtains $K_{01}^{pade}$ with $b_1=-C_2$.
When taking also the three-loop into account, there are two non-trivial possibilities: $K_{02}^{pade}$
with $b_1=-C_2$, $b_2=C_2^2-C_3$ and $K_{11}^{pade}$ with $b_1=-C_3/C_2$ and $a_1=C_2-C_3/C_2$. The various
results for the \pade approximants are shown in Fig.~ \ref{fig:padeKernel}, together with the corresponding
loop contributions $K_L=C_L x^{L-1}$. Note that, to determine the $C_L$ it is in principle sufficient to
evaluate the loop integrals $P_{L-loop}(k)$ for a single (small enough) value of the momentum $k$.

The \pade ansatz replaces the divergent behavior of the loop series
by an integrand kernel that is dominated by momentum modes $k\sim 0.01-0.1\, h/$Mpc, and is therefore not
very sensitive to the UV regime of the power spectrum (cf. Fig.~\ref{fig:padeKernel}). In addition, the three \pade approximants all feature an
integrable singularity that originates from a simple root of the polynomial in the denominator. The position
of this pole indicates the momentum scale where the perturbative expansion eventually breaks down. For the three
approximants we find the pole at $x\equiv\sigma_l^2(k,z)\simeq 1.4$ for $K_{01}^{pade}$, at $x=0.85$ for $K_{02}^{pade}$
and at $x=0.68$ for $K_{11}^{pade}$. In terms of the corresponding momentum $k_{pole}(z)$, this means that the pole
shifts to smaller momenta for smaller redshift, as can be also observed in Fig.~\ref{fig:padeKernel}. For example,
for $K_{02}^{pade}$ one has $k_{pole}(z=0)=0.3\, h$/Mpc and $k_{pole}(z=3)=3.1\, h$/Mpc. 

It is important to note that the integral in Eq.~(\ref{eq:Psmallk}) is still well-defined when taking the
principal value integral, since the pole of the \pade approximant of the kernel is integrable. This remedies the divergent
behavior observed in Eq.~(\ref{eq:smallkScaling}). Another way to view this is that the summation over
$L$ in Eq.~(\ref{eq:kloop}) has to be performed inside the momentum integral as in Eq.~(\ref{eq:Psmallk}),
and should not be interchanged with the loop integration over $q$ as done in the usual SPT loop
expansion, cf. Eq.~(\ref{eq:smallk}). 

This suggests that the divergent behavior of the loop expansion
in the small-$k$ limit is indeed spurious, and originates from interchanging the sum over loops and
the integration over the (largest) loop momentum $q$. The \pade ansatz provides a possibility to avoid this
issue. Furthermore, it can be improved systematically by increasing the order $n$ and $m$. By matching the
coefficients of the Taylor expansion of $K(x)$ in $x$ in Eq.~(\ref{eq:kloop}) to $L$-loop accuracy,
one could go up to approximants of order $n+m\leq L-1$. 
In this sense, a sequence of \pade approximations with increasing order provides a well-defined,
systematic way to improve the accuracy even beyond the three-loop matching considered here (to increase the order
to $n+m=3$ one would need $C_4$, which would require a four-loop computation).

One might wonder how much these findings change when introducing a finite UV cutoff $\Lambda_c$ in the loop integrals. As a first remark, we stress again that the loop integrals are finite and the cutoff used in our numerics corresponds to the $\Lambda_c\to\infty$ limit. Nevertheless, when imposing a (much) smaller cutoff, the results depend on $\Lambda_c$, as expected. We checked that, as long as $\Lambda_c \gg 1\,h/$Mpc, the dominant effect of the cutoff on the one-, two- and three-loop result is captured by replacing the upper integration limit in Eq.~(\ref{eq:smallk}) by $\Lambda_c$. This implies that the corresponding coefficients $C_L$ depend only relatively weakly on the cutoff. For example, $C_2=0.71\,(0.70)$ and $C_3=1.02\,(0.97)$ for $\Lambda_c=5\,(1)\,h/$Mpc, which is to be compared to $C_2=0.71$ and $C_3=1.05$ in the $\Lambda_c\to\infty$ limit. This implies that also the \pade approximant $K_{nm}$ is only mildly cutoff dependent. Since the integrand in Eq.~\eqref{eq:Psmallk} is dominated by modes $q\ll 1\,h/$Mpc (see Fig.~\ref{fig:padeKernel}), this means that also the \pade-resummed result for the power spectrum is rather robust with respect to imposing a finite cutoff $\Lambda_c$.

\subsection{\pade resummation for the power spectrum}

\begin{figure}[t]
\begin{center}
\includegraphics[width=0.7\textwidth]{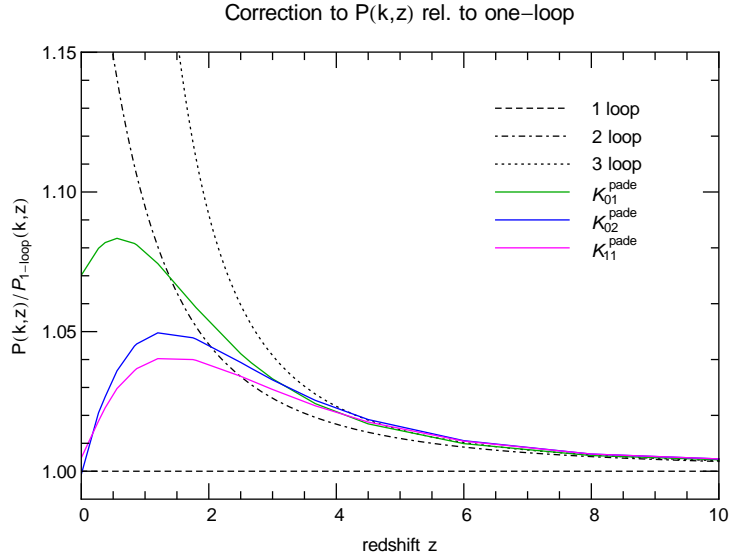}
\end{center}
\caption{\label{fig:pade} Correction to the power spectrum $P(k,z)$ at small $k$ vs. the redshift $z$. The $y$-axis is normalized to the small-$k$ limit of the one-loop correction $P^{small-k}_{1-loop}\equiv -\frac{61}{105} k^2\sigma_d^2 P_{lin}(k,z)$. The black lines correspond to the SPT result (two-loop dot-dashed, three-loop dotted), and the solid lines show the \pade resummed result when taking the one- and two-loop into account (green), and when taking also the three-loop into account (blue and magenta). Note that the plotted ratio is independent of $k$ when $k$ is small.}
\end{figure}

\begin{figure}[t]
\begin{center}
\includegraphics[width=0.45\textwidth]{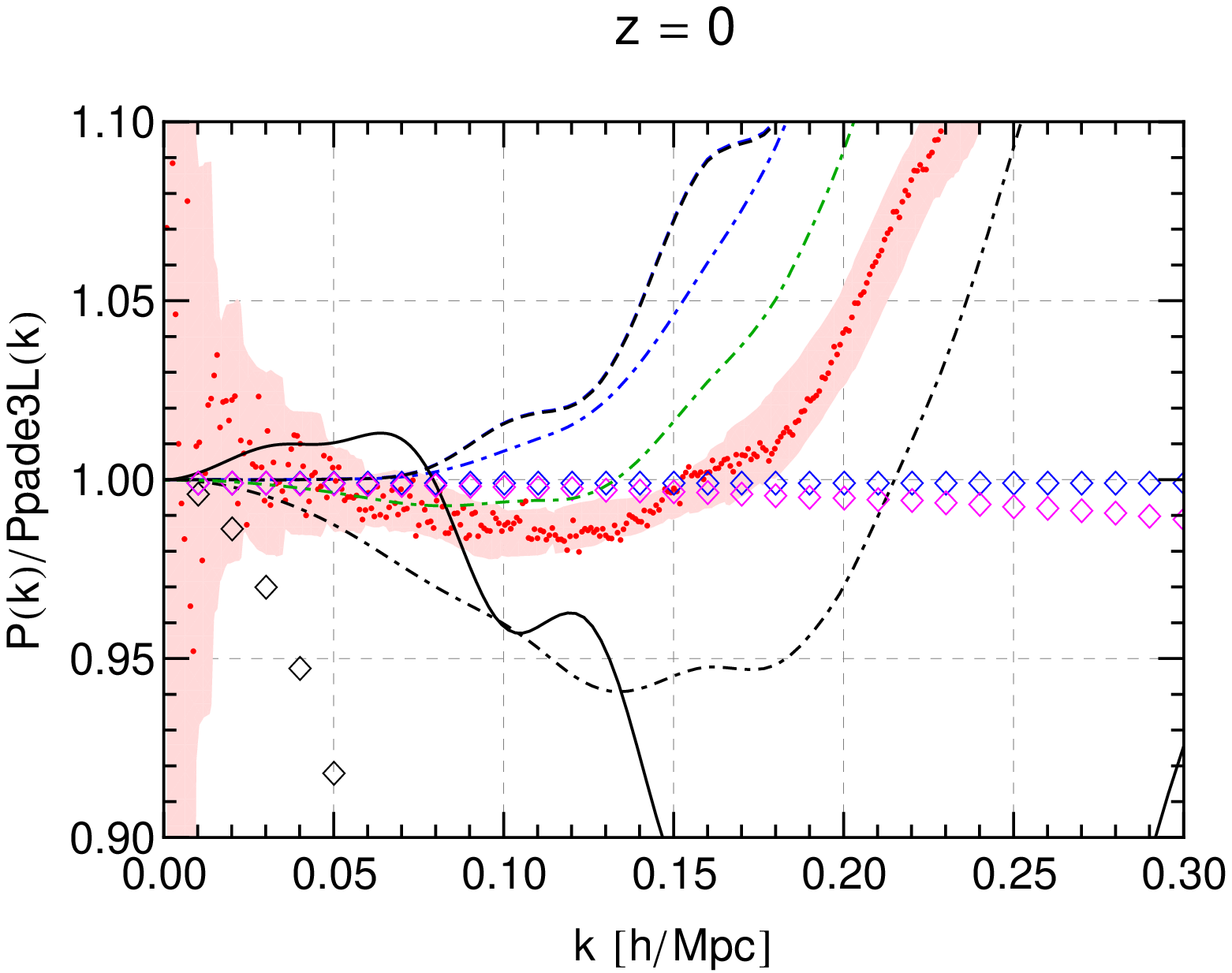}
\includegraphics[width=0.45\textwidth]{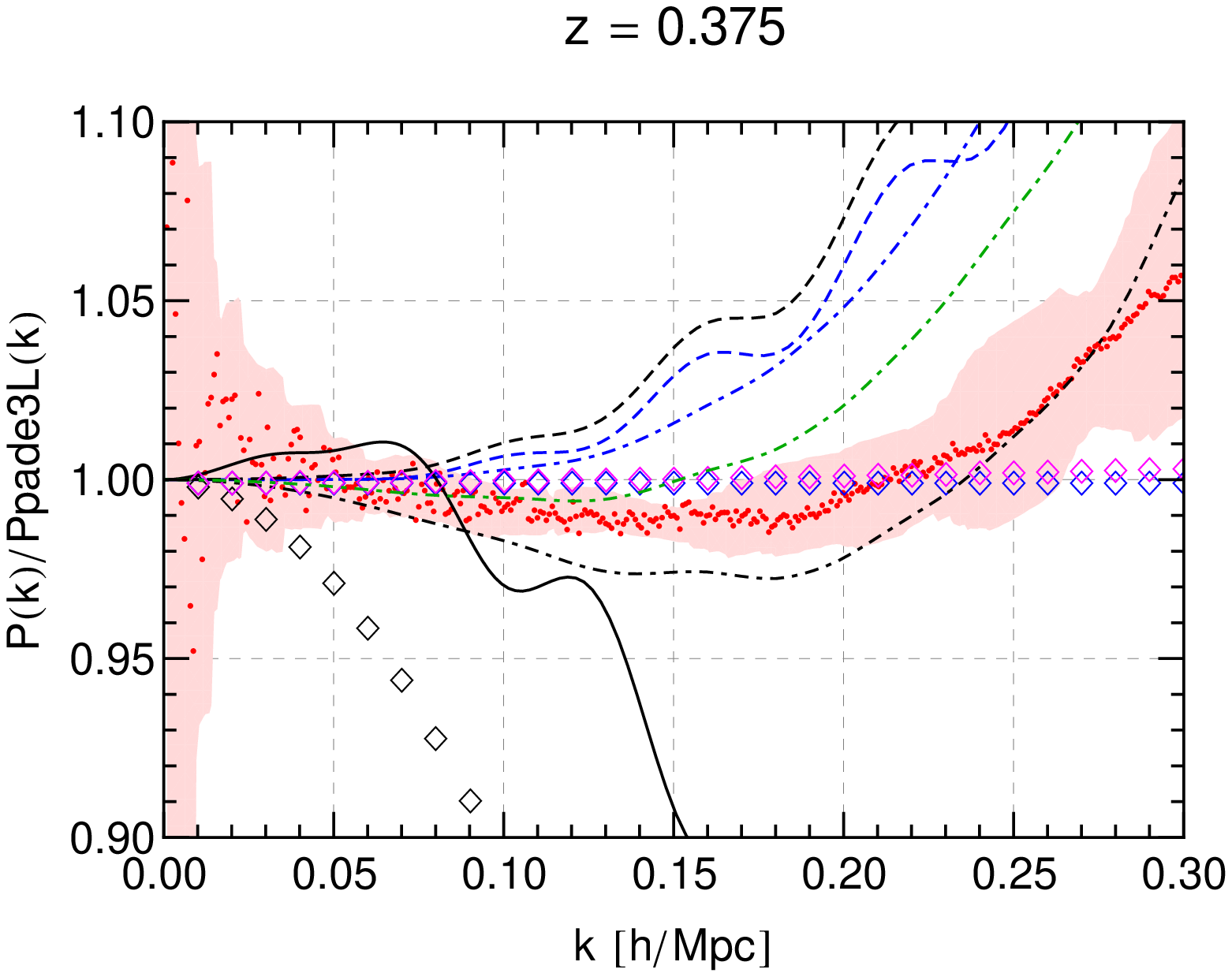}
\\
\includegraphics[width=0.45\textwidth]{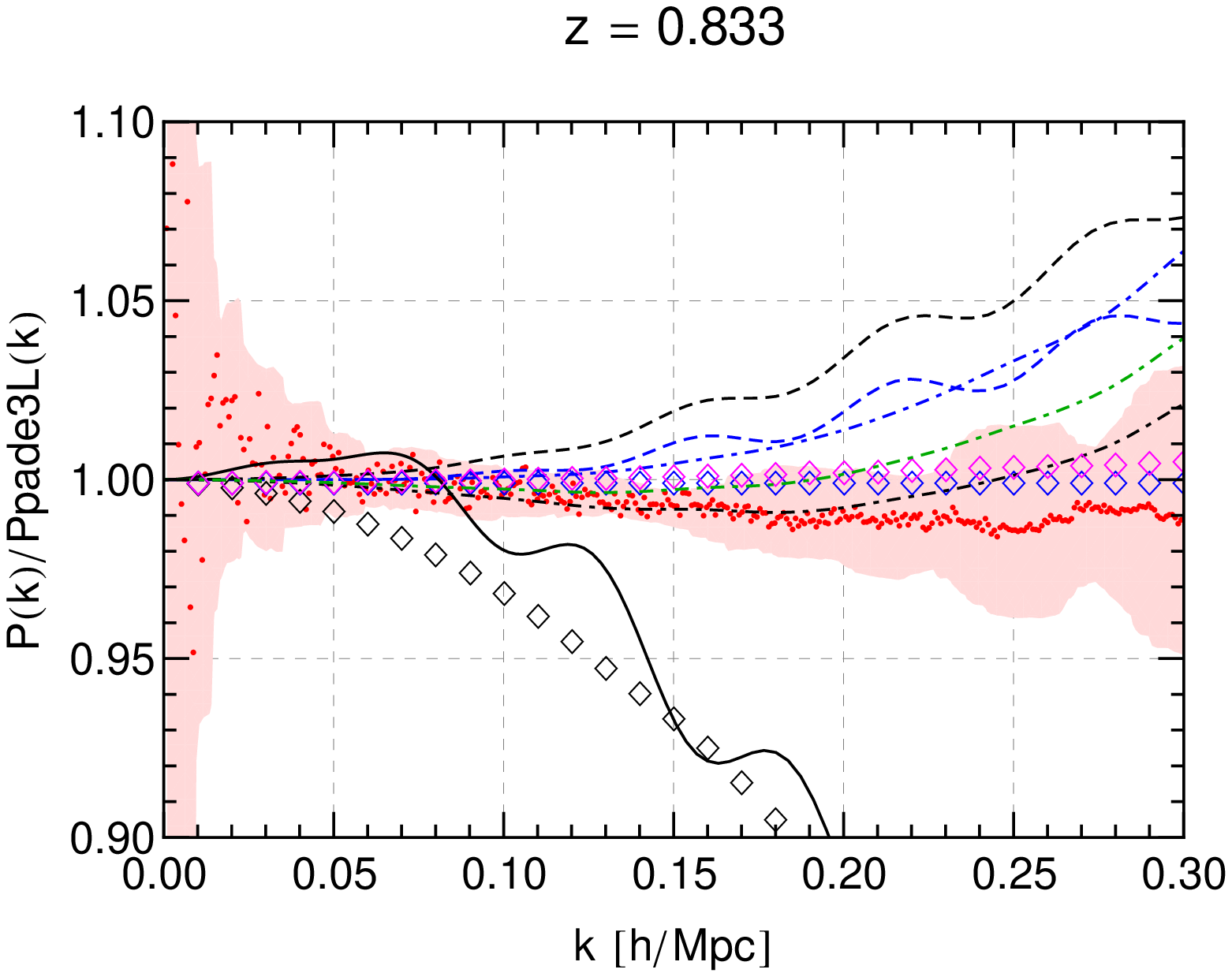}
\includegraphics[width=0.45\textwidth]{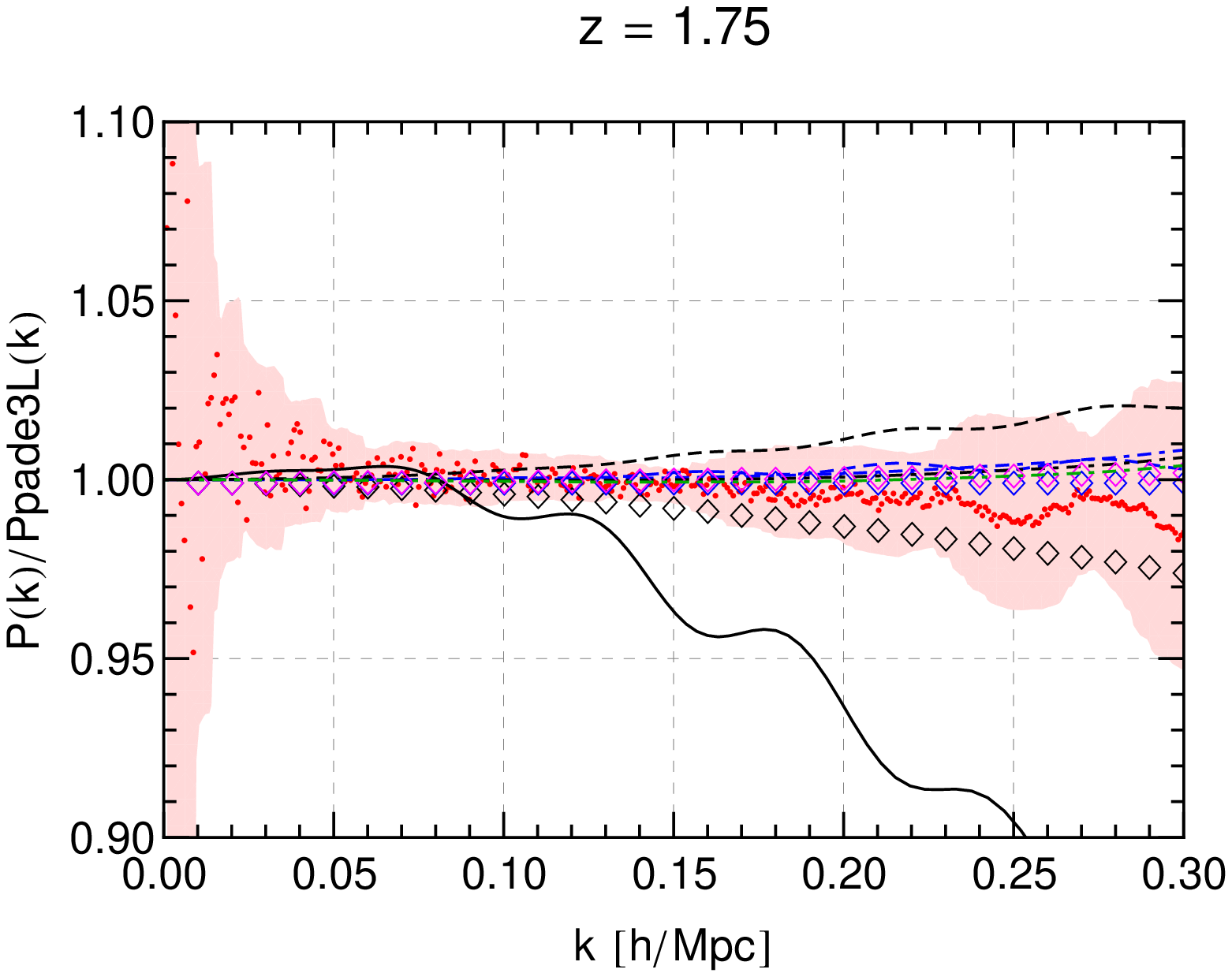}
\end{center}
\caption{\label{fig:padeNbody} As Fig.~\ref{fig:nbody_lowz}, but normalized to the \pade improved three-loop result with kernel $K^{pade}_{02}$ (blue diamonds). For comparison we show in addition \pade improved three-loop using $K^{pade}_{11}$ (magenta diamonds), and the \pade improved two-loop with the kernel $K^{pade}_{01}$ (green dotdashed line).}
\end{figure}

The results obtained for the small-$k$ limit of the power spectrum when inserting the various \pade approximants for the kernel $K$
in Eq.~(\ref{eq:Psmallk}) are shown in Fig.~\ref{fig:pade} as a function  of the redshift. Also shown are the perturbative two- and three-loop
contributions, whose divergent behavior for small $z$ can be clearly seen. In contrast, the results employing \pade approximants are well-behaved
even for $z\to 0$. It is reassuring to observe that the two approximants $K_{02}^{pade}$ and $K_{11}^{pade}$ yield very similar results, so that this ambiguity has only little effect on the final result. In addition, also the result obtained from the approximant $K_{01}^{pade}$, which is determined by matching only to two-loop order, is in reasonable agreement. Another observation is that, for large redshifts $z\gtrsim 3$ the \pade-resummed result agrees well with the perturbative three-loop result. For smaller redshifts $z\lesssim 2$, where the three-loop correction becomes larger than the two-loop contribution, the \pade resummed results are close to the two-loop value, and for $z\lesssim 1$ close to one-loop. This behavior is consistent with the one expected for an asymptotic series.

One can also extend this resummation to obtain an improved convergence behavior at momenta (slightly) above the regime where the
small-$k$ limit is strictly applicable. For that purpose, we consider a modified loop expansion
\bea
P(k,z) & = & P_{lin}(k,z) + P_{small-k}^{pade}(k,z) \nn\\
       && {} + P^{sub}_{1-loop}(k,z)  + P^{sub}_{2-loop}(k,z)  + P^{sub}_{3-loop}(k,z) + \dots \,,
\eea
where the \pade term is obtained by inserting a \pade approximant for the kernel $K^{pade}_{nm}$ in Eq.~(\ref{eq:Psmallk}), and
\bea
P^{sub}_{L-loop}(k,z) \equiv P_{L-loop}(k,z) - P^{small-k}_{L-loop}(k,z),
\eea
is the $L$-loop contribution in SPT with the small-$k$ limit subtracted. The latter is obtained by
inserting $K_L=C_L x^{L-1}$ for the kernel in Eq.~(\ref{eq:Psmallk}). 

In Figs.~\ref{fig:nbody_lowz} and \ref{fig:nbody_highz}, we show the results obtained using the \pade kernel $K^{pade}_{02}$ together with the subtracted SPT contributions up to one-, two- and three-loop, respectively (blue dashed and dot-dashed lines and blue diamonds -- we refer to them as \emph{\pade improved} one-, two- and three-loop results in the following).
We observe that the first three orders of this modified loop expansion exhibit a behavior that is much less divergent than without the \pade resummation. The improvement in convergence can be seen in some more detail in Fig~\ref{fig:padeNbody}, where we normalize the $y$-axis to
the \pade improved three-loop result and show a $\pm 10\%$ range. Evidently, the \pade improved loop expansion has significantly better
convergence properties (blue lines/diamonds) than SPT (black lines/diamonds).
\begin{table}[t]
  \centering
  \begin{tabular}{|c  | c |}
  \hline
  $z$ & $k_{max} / ( h / {\rm Mpc} )$ \\ 
  \hline
  $0$ & $0.11$ \\
  $0.375$ & $0.14$  \\
  $0.833$ & $0.18$  \\
  $1.75$ & $0.34$  \\
  \hline
  \end{tabular}
  \caption{The momentum $k_{max}$ denotes the scale where the \pade improved three-loop and two-loop results agree at the percent level, depending on the redshift.}
  \label{tab:kmax}
\end{table}
In particular, the difference between \pade improved three- and two-loop results is less than one percent for $k\lesssim 0.11\, h$ /Mpc at $z=0$. (More information is given in Table~\ref{tab:kmax}). In addition, in this range the prediction also agrees
with the N-body results. For the three-loop, the agreement with N-body is good for even somewhat larger momenta, depending on the redshift. However, for relatively large momentum $k\gtrsim 0.2\, h/$Mpc, the perturbative series still breaks down at $z=0$. Nevertheless, one may observe
that the simple \pade ansatz yields a considerable improvement compared to standard SPT even for momenta above the small-$k$ limit.

Finally, we would like to discuss the robustness of the \pade ansatz. In Fig.~\ref{fig:padeNbody} we show the results obtained when using the \pade kernels $K_{02}^{pade}$ or $K_{11}^{pade}$, respectively, together with the subtracted three-loop contribution from SPT
(blue and magenta diamonds, respectively). Their relative deviation is below one percent for $k\lesssim 0.32\, h/$Mpc at $z=0$, and is therefore negligible in practice. In addition, we also present a result that is obtained using the kernel $K_{01}^{pade}$ together with the subtracted two-loop contribution (green dot-dashed line in Fig.~\ref{fig:padeNbody}). Since this kernel relies on two-loop matching only, the corresponding result can be obtained without performing any three-loop calculation. It is interesting to observe that this result is very close to the \pade improved three-loop result with kernels $K_{02}^{pade}$ or $K_{11}^{pade}$ determined by three-loop matching even at $z=0$ for momenta up to $k\lesssim 0.15\, h/$Mpc. 

\section{Conclusions\label{sec:concl}}
 
There is currently a large interest in understanding the evolution of primordial density 
perturbations in the Universe beyond the linear predictions. 
This effort is driven both by the possibility of understanding physics related to the acceleration of the Universe and
of the medium that collapses to create the structure at cosmological scales 
\cite{Bernardeau:2001qr, Carrasco:2012cv, Hertzberg:2012qn,Amendola:2012ys, Baldi:2012ky}. 
Maybe more fundamentally, it is also spurred by the question if the perturbative expansions are consistent.

In the present work, we have investigated this issue by first numerically  determining the three-loop contributions to the power spectrum in SPT. 
We observed that the SPT series does not converge even for low momenta. 
As we explained, this is expected for a realistic spectrum. The series exhibits a behavior compatible with an asymptotic series. We exploited this to produce a \pade resummation that yields a significant improvement in the convergence of SPT. Quite interestingly, even at low-redshift this is not only relevant at low-$k$ but extends to scales 
 in the BAO momentum regime. Our approach is based on a \pade ansatz for a single function $K(x)$ of the variance of the density field $x=\sigma_l^2(k,z)$, whose parameters are determined by matching to the SPT $L$-loop corrections to the power spectrum at small $k$ (i.e. small $x$), $K(x)\sim\sum C_L x^{L-1}$. We stress that the coefficients $C_L$ are independent of momentum or redshift. Therefore, in principle, the matching coefficients can be obtained by evaluating the three-loop integrals only for a single momentum value. Additionally, the redshift dependence enters only via $x=\sigma_l^2(k,z)$. Hence, it may be regarded as a non-trivial check that the \pade ansatz improves the agreement with N-body data for all considered redshifts.

The features discussed above indicate that the improvement based on the \pade ansatz is systematic and not accidental. Also, it contains no free parameters and therefore the improvement is not biased. The comparison with N-body simulation supports the  claim that the \pade ansatz grasps some of the non-linear dynamics of the system in a wide range of scales. Nevertheless, the good agreement between the \pade resummation and the N-body results at $z \sim 0$ should not be overrated. Our main result is that one can achieve a convergent series using \pade approximants in the regime of small momenta. Notice that this distinguishes our resummation scheme 
from the ones that resum soft effects for large momenta. 
These typically reproduce SPT for low momenta and hence cannot converge.

There are different aspects of our approach that we leave for further research. First, it would be interesting to perform a cross-check using N-body simulations for various sets of parameters as well as a more thorough understanding of the behavior of the $C_L$ for large $L$.  Also, allowing for simulations with different
initial power spectra would be helpful to clarify the validity of our approach. Finally, it would be interesting to investigate
whether our \pade ansatz is related to approaches discussed in Refs. \cite{Tatekawa:2006gx,Matsubara:1997bx} to understand the
non-linear evolution of spheroidal bodies.

\section*{Acknowledgments}

We thank Ra\'ul Angulo, H\'ector Gil-Mar\'in, Juhan Kim and Rom\'an Scoccimarro  for very useful discussions. 
This work has been partially supported by the German Science Foundation (DFG) within the Collaborative Research Center 676 ``Particles, Strings and the Early Universe''.

\begin{appendix}

\section{Numerical integration \label{app:numerics}}

For the numerical evaluation of the loop integrals we use the Suave routine of the
Monte Carlo integration library CUBA 3.0 \cite{Hahn:2004fe}. We choose the external momentum $k$ along
the $z$-axis. The freedom to rotate all loop vectors by a common angle within the $x-y$
plane can be used to eliminate one polar integration. Furthermore, since the integrand is
symmetric (by construction) with respect to sign-flips of each of the loop momenta, it is
sufficient to integrate the azimuthal angles in the range $\cos\theta\in[0,1]$. At three loops,
we integrate over the eight-dimensional unit cube with $x_1-x_3$ chosen as in Eq.~(\ref{eq:absmom}), $x_4=\cos\theta_{k_1}$,
$x_5=\cos\theta_{k_2}$, $x_6=\cos\theta_{k_3}$, $x_7=\phi_{k_1}/(2\pi)$, $x_8=\phi_{k_2}/(2\pi)$, and set
$\phi_{k_3}=0$. We take the Jacobian for this substitution into account in the integrand.
Furthermore, we use $k_{min}=10^{-5}\,h/$Mpc, $k_{max}=10^2\, h/$Mpc, and
the Suave settings $\epsilon_{rel}=10^{-3}$, $\epsilon_{abs}=10^{-12}$,
$N_{eval}^{max}=10^8$, NOSMOOTH$=1$, LAST$=1$, NNEW$=5000$, FLATNESS=$25$. We checked that our results are stable
against variations in these parameters, and also checked that the routine Cuhre yields identical results within
the error estimate provided by CUBA. We checked
agreement of the two-loop results with the SPT output of the RegPT code  \cite{Taruya:2012ut}.

Ultimately, we used sequential logarithmic distributions for the absolute values of the loop momenta
\bea\label{eq:absmom}
\Q_1 &=& k_{min} \exp [ \log ( k_{max}/k_{min} ) x_1 ] \, , \nn  \\  
\Q_2 &=& k_{min} \exp [ \log ( \Q_1 /k_{min} )  x_2 ] \, , \nn  \\
\Q_3 &=& k_{min} \exp [ \log ( \Q_2 /k_{min} ) x_3 ] \, ,  
\eea
where $k_{min}$ and $k_{max}$ denote the minimal and maximal values of the input power spectrum and 
the variables $x_i$ are integrated over $[0,1]$. The reason for this choice is motivated by the asymptotic behavior in 
Eqs.~(\ref{eq:NLFS}) and (\ref{eq:largek}).

For Figs.~\ref{fig:power3L} and \ref{fig:power3LNorm} we used as input the linear spectrum obtained from CAMB for the WMAP5 parameters
$\Omega_m=0.279$, $\Omega_b/\Omega_m=0.165$, $n_s=0.96$, $h=0.701$, $\sigma_8=0.817$ (identical to \cite{Taruya:2012ut}). 
For all other figures, we used the set of $\Lambda$CDM parameters 
underlying the N-body simulation Horizon Run 2 as specified in \cite{Kim:2011ab}, which slightly differs from the previous ones.

As a cross-check, we also used an alternative parametrization for the absolute loop momenta given by
\be\label{eq:absmom2}
\Q_1 = k_{ref}\frac{x_1}{1-x_1} \, , \quad  
\Q_2 = \Q_1 x_2 \, , \quad
\Q_3 = \Q_2 x_3 \, ,  
\ee
with the external momentum as reference point, $k_{ref}=k$ . The results are in very good agreement with the ones obtained using Eq.~(\ref{eq:absmom}),
as can be seen by comparing the black diamonds with the grey crosses in Fig.~\ref{fig:power3L}. Finally, we
also checked that after the adequate modification of the kernels $F_n\to F_n^{ZA}$ \cite{Bernardeau:2001qr},
we reproduce the results to three-loops for the
ZA shown in \cite{Crocce:2005xy}.

\section{\pade resummation \label{app:pade}}

Let us first investigate the origin of Eq.~(\ref{eq:smallk}) more closely. The dominant $L$-loop contribution to the
power spectrum at small $k$ is given by \cite{Fry:1993bj,Valageas:2001td}
\begin{eqnarray}
 P_{L-loop}(k,z) &\to& \frac{(2L+1)!}{2^{L-1}L!}\, P_{lin}(k,z) \nonumber \\
&& {} \times \int_{q_1}\cdots\int_{q_L} F_{2L+1}^s(\vec
k,\vec q_1,-\vec q_1,\dots,\vec q_L,-\vec q_L) \nonumber\\
&& {} \times P_{lin}(q_1,z)\cdots P_{lin}(q_L,z) \;.
\label{eq:LowkP}
\end{eqnarray}
where $\int_q\equiv\int d^3q$.
Since $F_{2L+1}^s\propto k^2$ in the limit of small $k$ \cite{Goroff:1986ep,Bernardeau:2001qr}, the upper
term scales as $k^2P_{lin}(k,z)$. At one loop (see e.g.~\cite{Blas:2013bpa}),
\be
P_{1-loop}(k,z) \to -\frac{61}{105} k^2 P_{lin}(k,z) \frac{4\pi}{3}\int_0^\infty dq P_{lin}(q,z) = -\frac{61}{105} k^2\sigma_d^2 P_{lin}(k,z)\;,
\ee
for small $k$. Analogously, we define $L$-loop kernels ${\cal K}_L(q,z)$ by
\be
\label{eq:kernelLoop}
P_{L-loop}(k,z) \to  -\frac{61}{105} k^2 P_{lin}(k,z) \frac{4\pi}{3} \int_0^\infty dq P_{lin}(q,z) \, {\cal K}_L(q,z)\,.
\ee
By requiring that $q\equiv |\vec q_{max}|$ corresponds to the loop momentum with the largest absolute momentum, this definition is unique. The normalization is chosen such that ${\cal K}_1=1$. In general for $L\geq 2$
\bea
\frac{{\cal K}_L(|\vec q|,z)}{q^2} &=& - \frac{(2L+1)!}{2^{L-1}L!}\frac{315}{61}\, L \, \int_{q_2}\cdots\int_{q_L} {\cal F}_{2L+1}^s(\vec q,-\vec q,\vec q_2,-\vec q_2\dots,\vec q_L,-\vec q_L) \nonumber\\
&& {} \times P_{lin}(q_2,z)\cdots P_{lin}(q_L,z) \times \Theta(|\vec q|-|\vec q_2|)\cdots\Theta(|\vec q|-|\vec q_L|)\;,\nn
\eea
where ${\cal F}_{2L+1}^s(\vec q_1,-\vec q_1,\dots,\vec q_L,-\vec q_L) \equiv \lim_{k\to 0} F_{2L+1}^s(\vec
k,\vec q_1,-\vec q_1,\dots,\vec q_L,-\vec q_L)/k^2$ is the leading $k^2$ coefficient in the small-$k$ expansion
and the Heaviside functions as well as the prefactor $L$ account for choosing $\vec q_{max} = \vec q_1 (\equiv \vec q)$.
This is possible because $F_{2L+1}^s$ is totally symmetric in the $\vec q_i$. Note that, due to rotational symmetry,
${\cal K}_L$ depends only on the absolute value of $\vec q$.

The coefficients used for the matching in the \pade ansatz are then given by
\be
C_L = \frac{\int_0^\infty dq P_{lin}(q,z) \, {\cal K}_L(q,z)}{\int_0^\infty dq P_{lin}(q,z) \, \sigma_l^{2L-2}(q,z)}\;.
\ee
Note that the redshift-dependence cancels such that the left-hand side is indeed independent of $z$, and that $C_1=1$.
The $C_L$ are also independent of the external momentum $k$ by construction.
In the momentum regime $k\ll |\vec q_i| \ll q$ for $i=2,\dots,L$ one has $F^s_{2L+1}\propto k^2/q^2$, i.e. ${\cal F}^s_{2L+1}\propto 1/q^2$.
If this proportionality were exact, then even the dependence of $C_L$ on $\sigma_l$ (or equivalently on the shape of $P_{lin}$) would drop
out completely and the $C_L$ would be constants, independent of the input power spectrum. Deviations from this
simple scaling for $|\vec q_i| \lesssim q$ lead to a sensitivity on $P_{lin}$, which is moderate at two-loops, see below.

At two-loop one finds \cite{Blas:2013bpa} (see also \cite{Bernardeau:2012ux})
\be
{\cal K}_2(q,z) = \frac{44764}{83265} 4\pi \int_0^q\, dp\, p^2\, g(p/q) P_{lin}(p,z)
\ee
where
\begin{eqnarray}
 g(x) &=& \frac{1}{179056
x^6}\Bigg((x^2+1)\left(128258x^4-5760(x^8+1)-13605(x^6+x^2)\right)\nonumber\\
 && {} -
\frac{15}{4x}(x^2-1)^4\left(384(x^4+1)+2699x^2\right)\ln\left(\frac{x-1}{x+1}
\right)^2 \Bigg)\;.
\end{eqnarray}

Using that $1\leq g(x)\leq 120424/78337\simeq 1.54$ in the relevant range $1\geq x\geq 0$, one obtains the strict inequality
\be
 0.54 \simeq \frac{44764}{83265} \leq C_2 \leq \frac{44764}{83265}\times \frac{120424}{78337} \simeq 0.83 \;,
\ee
which is valid for an arbitrary choice of $P_{lin}$.
Note that the value $C_2\simeq 0.71$ obtained for a linear spectrum corresponding to WMAP5 parameters lies
close to the middle of this interval. 

\section{N-body data \label{app:hr3}}

\begin{figure}[t]
\begin{center}
\includegraphics[width=0.45\textwidth]{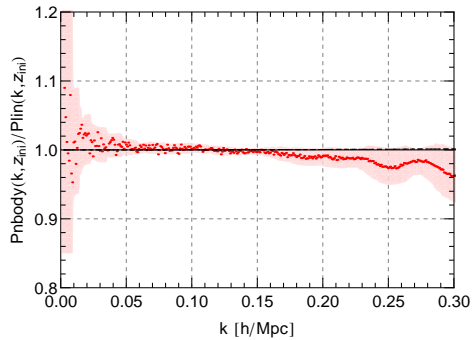}
\end{center}
\caption{\label{fig:nbodyini} Power spectrum as provided by the
Horizon Run 2 data \cite{Kim:2011ab} at the initialization redshift $z_{ini}=32$
divided by the linear spectrum from CAMB. The shaded region corresponds to the
estimate of the relative uncertainty in the N-body data used in Figs.~\ref{fig:nbody_lowz},~\ref{fig:nbody_highz} and \ref{fig:padeNbody}.}
\end{figure}

We use the N-body data of the Horizon Run 2 presented in \cite{Kim:2011ab} as comparison with our analytical methods. This data is based on a box of size $7200\, {\rm Mpc}/h$
and $N=6000^3$ particles with mean separation $1.2\, \mathrm{Mpc}/h$, initialized at $z_{ini}=32$.
The power spectrum at $z_{ini}$ shows slight deviations from the linear spectrum due
to sampling variance at low $k$ and is slightly suppressed at larger $k$ due to a convolution 
with the density assignment kernel, see Fig.~\ref{fig:nbodyini}. Although it is non-trivial
how this deviation propagates to smaller $z$, we use the deviation $R(k)\equiv P_{n-body}(k,z_{ini})/P_{lin}(k,z_{ini})$ as an estimate
to bracket the relative uncertainty of $P_{n-body}(k,z)$. The shaded areas shown in Figs.~\ref{fig:nbody_lowz},~\ref{fig:nbody_highz},~\ref{fig:padeNbody} and \ref{fig:nbodyini} correspond to the intervals $[$min$(R(k),R(k)^{-1})P_{n-body}(k,z),$ max$(R(k),R(k)^{-1})P_{n-body}(k,z)]$. 

We checked that using the larger simulation Horizon Run 3 yields very similar results. However, it has a slightly worse resolution
at small scales, which would lead to larger uncertainties (i.e. larger max$(R(k),R(k)^{-1})$) at $k\gtrsim 0.15\, h/$Mpc (by a factor $1.5-2$).

\end{appendix}



\begin{thebibliography}{99}

\bibitem{Crocce:2005xy}
  M.~Crocce and R.~Scoccimarro,
  ``Renormalized cosmological perturbation theory,''
  Phys.\ Rev.\ D {\bf 73} (2006) 063519
  [astro-ph/0509418].

\bibitem{Crocce:2005xz}
  M.~Crocce and R.~Scoccimarro,
  ``Memory of initial conditions in gravitational clustering,''
  Phys.\ Rev.\ D {\bf 73} (2006) 063520
  [astro-ph/0509419].

\bibitem{Nishimichi:2008ry}
  T.~Nishimichi, A.~Shirata, A.~Taruya, K.~Yahata, S.~Saito, Y.~Suto, R.~Takahashi and N.~Yoshida {\it et al.},
  ``Modeling Nonlinear Evolution of Baryon Acoustic Oscillations: Convergence Regime of N-body Simulations and Analytic Models,'' Publications of the Astronomical Society of Japan, {\bf 61}, No.2, 321
  [arXiv:0810.0813 [astro-ph]]. 

\bibitem{Carlson:2009it}
  J.~Carlson, M.~White and N.~Padmanabhan,
  ``A critical look at cosmological perturbation theory techniques,''
  Phys.\ Rev.\ D {\bf 80} (2009) 043531
  [arXiv:0905.0479 [astro-ph.CO]].

\bibitem{Scoccimarro:1995if}
  R.~Scoccimarro and J.~Frieman,
  ``Loop corrections in nonlinear cosmological perturbation theory,''
  Astrophys.\ J.\ Suppl.\  {\bf 105} (1996) 37
  [astro-ph/9509047].

\bibitem{Kehagias:2013yd}
  A.~Kehagias and A.~Riotto,
  ``Symmetries and Consistency Relations in the Large Scale Structure of the Universe,''
  Nucl.\ Phys.\ B {\bf 873} (2013) 514
  [arXiv:1302.0130 [astro-ph.CO]].

\bibitem{Peloso:2013zw}
  M.~Peloso and M.~Pietroni,
  ``Galilean invariance and the consistency relation for the nonlinear squeezed bispectrum of large scale structure,''
  JCAP {\bf 1305} (2013) 031
  [arXiv:1302.0223 [astro-ph.CO]].

\bibitem{Blas:2013bpa}
  D.~Blas, M.~Garny and T.~Konstandin,
  ``On the non-linear scale of cosmological perturbation theory,''
  arXiv:1304.1546 [astro-ph.CO].
  
\bibitem{Anselmi:2012cn}
  S.~Anselmi and M.~Pietroni,
  ``Nonlinear Power Spectrum from Resummed Perturbation Theory: a Leap Beyond the BAO Scale,''
  JCAP {\bf 1212} (2012) 013
  [arXiv:1205.2235 [astro-ph.CO]].

\bibitem{Sugiyama:2013pwa}
  N.~S.~Sugiyama and T.~Futamase,
  ``Relation between standard perturbation theory and regularized multi-point propagator method,''
  Astrophys.\ J.\  {\bf 769} (2013) 106
  [arXiv:1303.2748 [astro-ph.CO]].

\bibitem{Sugiyama:2013gza}
  N.~S.~Sugiyama and D.~N.~Spergel,
  ``How does non-linear dynamics affect the baryon acoustic oscillation?,''
  arXiv:1306.6660 [astro-ph.CO].

\bibitem{Carrasco:2013sva}
  J.~J.~M.~Carrasco, S.~Foreman, D.~Green and L.~Senatore,
  ``The 2-loop matter power spectrum and the IR-safe integrand,''
  arXiv:1304.4946 [astro-ph.CO].

\bibitem{Taruya:2012ut}
  A.~Taruya, F.~Bernardeau, T.~Nishimichi and S.~Codis,
  ``RegPT: Direct and fast calculation of regularized cosmological power spectrum at two-loop order,''
  Phys.\ Rev.\ D {\bf 86} (2012) 103528
  [arXiv:1208.1191 [astro-ph.CO]].

\bibitem{Bernardeau:2001qr}
  F.~Bernardeau, S.~Colombi, E.~Gaztanaga and R.~Scoccimarro,
  ``Large scale structure of the universe and cosmological perturbation theory,''
  Phys.\ Rept.\  {\bf 367} (2002) 1
  [astro-ph/0112551].

\bibitem{Valageas:2013hxa}
  P.~Valageas,
  ``Accuracy of analytical models of the large-scale matter distribution,''
  arXiv:1308.6755 [astro-ph.CO].


\bibitem{Carrasco:2012cv}
  J.~J.~M.~Carrasco, M.~P.~Hertzberg and L.~Senatore,
  ``The Effective Field Theory of Cosmological Large Scale Structures,''
  JHEP {\bf 1209} (2012) 082
  [arXiv:1206.2926 [astro-ph.CO]].

\bibitem{Hertzberg:2012qn}
  M.~P.~Hertzberg,
  ``The Effective Field Theory of Dark Matter and Structure Formation: Semi-Analytical Results,''
  arXiv:1208.0839 [astro-ph.CO].


\bibitem{Peebles:Book}
P.~J.~E.~Peebles, ``The Large-Scale Structure of the Universe'' (Princeton
University Press, 1980)

\bibitem{Pueblas:2008uv}
  S.~Pueblas and R.~Scoccimarro,
  ``Generation of Vorticity and Velocity Dispersion by Orbit Crossing,''
  Phys.\ Rev.\ D {\bf 80} (2009) 043504
  [arXiv:0809.4606 [astro-ph]].

\bibitem{Pietroni:2011iz}
  M.~Pietroni, G.~Mangano, N.~Saviano and M.~Viel,
  ``Coarse-Grained Cosmological Perturbation Theory,''
  JCAP {\bf 1201} (2012) 019
  [arXiv:1108.5203 [astro-ph.CO]].

\bibitem{Lewis:1999bs}
  A.~Lewis, A.~Challinor and A.~Lasenby,
  ``Efficient computation of CMB anisotropies in closed FRW models,''
  Astrophys.\ J.\  {\bf 538} (2000) 473
  [astro-ph/9911177].

\bibitem{Goroff:1986ep}
  M.~H.~Goroff, B.~Grinstein, S.~J.~Rey and M.~B.~Wise,
  ``Coupling of Modes of Cosmological Mass Density Fluctuations,''
  Astrophys.\ J.\  {\bf 311} (1986) 6.

\bibitem{Hahn:2004fe}
  T.~Hahn,
  ``CUBA: A Library for multidimensional numerical integration,''
  Comput.\ Phys.\ Commun.\  {\bf 168} (2005) 78
  [hep-ph/0404043].
  
\bibitem{Eisenstein:1997jh}
  D.~J.~Eisenstein and W.~Hu,
  ``Power spectra for cold dark matter and its variants,''
  Astrophys.\ J.\  {\bf 511} (1997) 5
  [astro-ph/9710252].
  
\bibitem{Dyson:1952tj}
  F.~J.~Dyson,
  ``Divergence of perturbation theory in quantum electrodynamics,''
  Phys.\ Rev.\  {\bf 85} (1952) 631.

\bibitem{Bernardeau:2012ux}
  F.~Bernardeau, A.~Taruya and T.~Nishimichi,
  ``Cosmic propagators at two-loop order,''
 arXiv:1211.1571 [astro-ph.CO].

\bibitem{Bender:Book}
C.~M.~Bender and S.~A.~Orszag, ``Advanced mathematical methods for scientists and engineers'' (McGraw-Hill , 1978)

\bibitem{Kim:2011ab}
  J.~Kim, C.~Park, G.~Rossi, S.~M.~Lee and J.~R.~Gott, III,
  ``The New Horizon Run Cosmological N-Body Simulations,''
  J.\ Korean Astron.\ Soc.\  {\bf 44} (2011) 217
  [arXiv:1112.1754 [astro-ph.CO]].

\bibitem{Dunne:2004nc}
  G.~V.~Dunne,
  ``Heisenberg-Euler effective Lagrangians: Basics and extensions,''
  In *Shifman, M. (ed.) et al.: From fields to strings, vol. 1* 445-522
  [hep-th/0406216].

\bibitem{Moore:2000jw}
  G.~D.~Moore and K.~Rummukainen,
  ``Electroweak bubble nucleation, nonperturbatively,''
  Phys.\ Rev.\ D {\bf 63} (2001) 045002
  [hep-ph/0009132].

\bibitem{Tatekawa:2006gx}
  T.~Tatekawa,
  ``Improving of the Lagrangian perturbative solution for cosmic fluid: Applying Shanks transformation,''
  Phys.\ Rev.\ D {\bf 75} (2007) 044028
  [astro-ph/0605250].

\bibitem{Matsubara:1997bx}
  T.~Matsubara, A.~Yoshisato and M.~Morikawa,
  ``Beyond Zeldovich type approximations in gravitational instability theory: Pade prescription in spheroidal collapse,''
  Astrophys.\ J.\  {\bf 504} (1998) 7
  [astro-ph/9708154].

\bibitem{NadkarniGhosh:2010th}
  S.~Nadkarni-Ghosh and D.~F.~Chernoff,
  ``Extending the domain of validity of the Lagrangian approximation,''
  Mon.\ Not.\ Roy.\ Astron.\ Soc.\  {\bf 410} (2011) 1454
  [arXiv:1005.1217 [astro-ph.CO]].

\bibitem{Bernardeau:2008fa}
  F.~Bernardeau, M.~Crocce and R.~Scoccimarro,
  ``Multi-Point Propagators in Cosmological Gravitational Instability,''
  Phys.\ Rev.\ D {\bf 78} (2008) 103521
  [arXiv:0806.2334 [astro-ph]].

\bibitem{Okamura:2011nu}
  T.~Okamura, A.~Taruya and T.~Matsubara,
  ``Next-to-leading resummation of cosmological perturbations via the Lagrangian picture: 2-loop correction in real and redshift spaces,''
  JCAP {\bf 1108} (2011) 012
  [arXiv:1105.1491 [astro-ph.CO]].

\bibitem{Percival:2007yw}
  W.~J.~Percival, S.~Cole, D.~J.~Eisenstein, R.~C.~Nichol, J.~A.~Peacock, A.~C.~Pope and A.~S.~Szalay,
  ``Measuring the Baryon Acoustic Oscillation scale using the SDSS and 2dFGRS,''
  Mon.\ Not.\ Roy.\ Astron.\ Soc.\  {\bf 381} (2007) 1053
  [arXiv:0705.3323 [astro-ph]].

\bibitem{Pade:Book}
G.~A.~Baker Jr. (Ed.) and J.~L.~Gammel (Ed.), ``Pade Approximant in Theoretical Physics'' (Elsevier Science , 1970)

\bibitem{Amendola:2012ys}
  L.~Amendola {\it et al.}  [Euclid Theory Working Group Collaboration],
  ``Cosmology and fundamental physics with the Euclid satellite,''
  Living Rev.\  Relativity {\bf 16,} (2013) , 6
  [arXiv:1206.1225 [astro-ph.CO]].

\bibitem{Baldi:2012ky}
  M.~Baldi,
  ``Dark Energy Simulations,''
  Phys.\ Dark Univ.\  {\bf 1} (2012) 162
  [arXiv:1210.6650 [astro-ph.CO]].


\bibitem{Fry:1993bj}
  J.~N.~Fry,
  ``The Minimal power spectrum: Higher order contributions,''
  Astrophys.\ J.\  {\bf 421} (1994) 21.

\bibitem{Valageas:2001td}
  P.~Valageas,
  ``Dynamics of gravitational clustering v. subleading corrections in the quasi-linear regime,''
   A \& A, {\bf 382} (2002) 477 - 487,
   [arXiv:astro-ph/0109408].

\end{thebibliography}
\end{document}